\documentclass[10pt,twocolumn,superscriptaddress,aps,prb,balancelastpage,longbibliography]{revtex4-1}
\usepackage[latin9]{inputenc}
\usepackage{color,multirow}
\usepackage{verbatim}
\usepackage{amsmath}
\usepackage{mathtools}
\usepackage{amssymb,ulem,manfnt}
\usepackage{graphicx,cellspace,booktabs}
\usepackage{esint}
\usepackage[unicode=true,pdfusetitle,
 bookmarks=true,bookmarksnumbered=false,bookmarksopen=false,
 breaklinks=false,pdfborder={0 0 1},backref=false,colorlinks=true]{hyperref}
\usepackage{breakurl}
\usepackage{epstopdf}
\usepackage{yfonts}
\usepackage[dvipsnames]{xcolor}

\makeatletter

\@ifundefined{textcolor}{}
{%
 \definecolor{BLACK}{gray}{0}
 \definecolor{WHITE}{gray}{1}
 \definecolor{RED}{rgb}{1,0,0}
 \definecolor{GREEN}{rgb}{0,1,0}
 \definecolor{BLUE}{rgb}{0,0,1}
 \definecolor{CYAN}{cmyk}{1,0,0,0}
 \definecolor{MAGENTA}{cmyk}{0,1,0,0}
 \definecolor{YELLOW}{cmyk}{0,0,1,0}
}

\usepackage[caption=false]{subfig}
\usepackage{bm}

\newcommand{\bra}[1]{\langle #1 |}
\newcommand{\ket}[1]{|#1\rangle}

\def\l@subsubsection#1#2{}
\makeatother

\begin{document}
\title{Extraction of many-body Chern number from a single wave function}
\author{Hossein Dehghani}
\affiliation{Joint Quantum Institute, College Park, 20742 MD, USA}
\affiliation{The Institute for Research in Electronics and Applied Physics, University of Maryland, College Park, 20742 MD, USA}

\author{Ze-Pei Cian}
\affiliation{Joint Quantum Institute, College Park, 20742 MD, USA}

\author{Mohammad Hafezi}
\affiliation{Joint Quantum Institute, College Park, 20742 MD, USA}
\affiliation{The Institute for Research in Electronics and Applied Physics, University of Maryland, College Park, 20742 MD, USA}

\author{Maissam Barkeshli}
\affiliation{Condensed Matter Theory Center, Department of Physics, University of Maryland, College Park, 20742 MD, USA}
\affiliation{Joint Quantum Institute, College Park, 20742 MD, USA}

\begin{abstract}
  The quantized Hall conductivity of integer and fractional quantum Hall (IQH and FQH) states is directly related to a topological invariant, the many-body Chern number. The conventional calculation of this invariant in interacting systems requires a family of many-body wave functions parameterized by twist angles in order to calculate the Berry curvature. In this paper, we demonstrate how to extract the Chern number given a single many-body wave function, without knowledge of the Hamiltonian. 
  For FQH states, our method requires one additional integer invariant as input: the number of $2\pi$ flux quanta, $s$, that must be inserted to obtain a topologically trivial excitation. As we discuss, $s$ can be obtained in principle from the degenerate set of ground state wave functions on the torus, without knowledge of the Hamiltonian.
  We perform extensive numerical simulations involving IQH and FQH states to validate these methods.
\end{abstract}

\maketitle

\section{Introduction}

The integer and fractional quantized Hall conductivities of quantum Hall states provide some of the most well-known and striking experimental examples of topological quantization in physics. The existence of such quantized Hall conductivities is directly related to a $U(1)$ symmetry protected topological invariant of gapped many-body quantum states in two spatial dimensions, referred to as the many-body Chern number. Integer and fractional quantum Hall (IQH and FQH) states both possess a well-defined integer-valued Chern number. 

The conventional methods for defining the Chern number require a family of many-body ground states with twisted boundary conditions \cite{Niu1985Quantized, kohmoto1985topological}. For example, the Chern number is often defined in terms of the Berry curvature in the space of twisted boundary conditions of the many-body ground state on a torus. To calculate the Chern number with these methods thus requires access to a family of many-body ground states parameterized by the twist angle either in the entire twist angle parameter space or in a small region of it  \cite{hastings2015, Kudo2019Many}. This raises a fundamental conceptual question of whether it is possible to obtain the Chern number from a single ground state wave function -- or even simply a reduced density matrix on an open disk-like patch of the system -- and without knowledge of the Hamiltonian. Stated differently, is the Chern number even a well-defined property of a single ground state wave function? 

 Moreover, this issue is of practical interest for both numerical and experimental studies of topological states of matter, as the conventional methods include transport or magnetic fluxes piercing a hole of the system, to obtain a family of many-body ground states as a function of a continuous parameter. In recent years there has been substantial interest in engineering topological states of matter in various designer quantum systems, such as neutral atoms in optical lattices \cite{ Cooper2019Topological}, superconducting qubits \cite{ houck2012chip,roushan2017chiral}, photons \cite{ozawa2019topological} and potentially Rydberg arrays \cite{browaeys2020many}. The development of these platforms for realizing many-body topological states bring with it a necessity to find alternative methods to measure topological invariants such as the many-body Chern number, as these systems may not be amenable to standard transport measurements\cite{repellin2019detecting}.  Alternative approaches to measuring topological invariants require an ancilla to be coupled to the entire system and involve many-body Ramsey interferometry to measure the fractional charge \cite{grusdt2016interferometric}, entanglement entropy \cite{pichler2016measurement}, or modular matrices \cite{zhu2017}.

In this paper, we show how to compute the Chern number given a single ground state wave function. In particular, we demonstrate that the wave function on a cylindrical geometry or an open disk-like patch is sufficient to obtain the Chern number. For cylindrical geometries we provide both a topological quantum field theory explanation and numerical simulations to prove the validity of our formulas. In the case of disk-like geometries our analytical understanding is less complete although we provide extensive empirical numerical evidence that the same method is applicable. For both of these geometries, we present two families of formulas for computing the many-body Chern number given a single ground state wave function. While one family uses a single copy of the state, the other family uses two copies of the given state. We refer to these two families of formulas as the single layer and bilayer formulas, respectively. To use our results for experimentally detecting the many-body Chern number, the single layer formulas require many-body phase measurement techniques such as Ramsey interferometry, which is a costly measurement and requires introducing ancilla qubits to the system. In contrast, the two-copy formula can be combined with random local unitary measurements, as we have demonstrated in \cite{PRL}, to measure the Chern number without requiring many-body interferometry techniques.

For FQH states, our method requires one additional invariant, $s$, to obtain the integer Chern number $C$. $s$ is defined to be the number of flux quanta that must be adiabatically inserted into a region of the system before a topologically trivial excitation is obtained. For a state with Hall conductivity $\sigma_{H} = \frac{p}{q} \frac{e^2}{h}$, with $p$ and $q$ coprime, $s = r q$, for some integer $r$. The Chern number $C$ that we calculate then corresponds to $C = r p$. Thus, given $s$ and a single ground state wave function on a disk, we can obtain the fractional quantized Hall conductivity $\sigma_{H} = \frac{C}{s} \frac{e^2}{h}$. We discuss how $s$ can be obtained given the degenerate set of ground state wave functions of the system defined on a torus (without requiring their dependence on additional parameters such as twist angles). 

We emphasize that our results hold for strongly interacting many-body states. For disordered free fermion Hamiltonians, there are known methods to obtain the Chern number without twisting boundary conditions given the single-particle eigenstates \cite{hastings2010, kitaev2006anyons}, for example by computing a certain topological obstruction for almost commuting matrices \cite{hastings2010}.

From a broader perspective, topologically ordered phases of matter with symmetry, referred to as symmetry-enriched topologically ordered states (SETs), possess a host of topological invariants \cite{Barkeshli2019Symmetry}. Some of these topological invariants only depend on the intrinsic topological order, irrespective of the symmetry of the system, and are related to the properties of fusion and braiding of topologically non-trivial excitations. Other topological invariants, such as the many-body Chern number, require symmetry to define and are determined by the patterns of symmetry fractionalization and the fusion and braiding properties of symmetry defects.
Recently a general algebraic theory has been presented in terms of a G-crossed braided tensor category to completely characterize symmetry-enriched topological phases for arbitrary symmetry groups by characterizing the fusion and braiding properties of anyons and symmetry defects \cite{Barkeshli2019Symmetry}. It is a fundamental open question to understand how and to what extent, given a single ground state wave function or reduced density matrix on a local patch of space, the many-body topological invariants predicted by this theory can be extracted. 

A more trivial class of states, referred to as symmetry-protected topological (SPT) states \cite{Senthil2015SPT}, have no intrinsic topological order and are adiabatically connected to trivial product states if the symmetry is broken. For SPTs, there have been a series of works demonstrating how to extract certain symmetry-protected many-body topological invariants in various special cases \cite{denNijs1989preroughening, Pollmann2012symmetry, Cirac2012order, Shiozaki2017, Shiozaki2018many}, and how to measure them using random unitary operations \cite{elben2019many}. So far, these works have focused mainly on the case where the symmetry group is finite, where one can relate matrix elements of certain operators to certain finite group cohomology invariants. However for SET states, which have intrinsic topological order, the question of extracting symmetry-protected many-body topological invariants from a single many-body wave function has received little, if any, attention. 

We note that our approach is particularly inspired by results of Ref. \onlinecite{Shiozaki2018many}. In particular, one of the formulas we present is closely related to, but distinct from, a formula of Ref. \onlinecite{Shiozaki2018many} for extracting the Chern number for integer Chern insulators using the ground state on a cylinder. To obtain correct results, the formula of Ref. \onlinecite{Shiozaki2018many} needs to be modified, as we discuss below. 

\subsection*{Summary of results}

Let us consider a gapped many-body ground state $|0\rangle$ in (2+1)D with $U(1)$ symmetry. The system possesses an integer-valued invariant, the many-body Chern number $C$, which determines the Hall conductivity through the formula
\begin{align}
\sigma_H  = C \frac{1}{s} \frac{e^2}{h} = \frac{p}{q} \frac{e^2}{h}, 
\end{align}
where $p$ and $q$ are coprime. Here $s$ is another integer invariant that can be understood as follows. We consider the flux insertion operator $\hat{\Phi}_x$ for adiabatic insertion of $2\pi$ flux through the $x$ cycle of the torus. $s$ is then the minimal integer such that, given any ground state $|\psi\rangle$ on a torus, 
\begin{align}
\hat{\Phi}_x^s |\psi\rangle = |\psi \rangle. 
\end{align}
Alternatively, adiabatically inserting flux at some point in the system gives a Laughlin quasihole, $v$, which we refer to as the vison. $v$ is in general a topologically non-trivial excitation with fractional charge $Q_v = p/q$, and statistics $\theta_v = \pi p/q$. $s$ is defined as the minimal integer such that $v^s$ is a topologically trivial excitation. Furthermore, as we show in Appendix \ref{qpChargeS}, all fractional electric charges of the quasiparticles are integer multiples of $1/s$. For an IQH state which does not host any topologically non-trivial excitations, $s = q = 1$. However for a general FQH state, it follows that $s = r q$ for some integer $r$. 

The state $|0\rangle$ that we consider can be defined on any space. For concreteness we take it to be defined on either a cylindrical geometry or an open region with coordinates $(x,y) \in [0, L_x] \times [0,L_y]$. In the cylindrical case, we take the $y$ direction to be compactified. 

We present the following formulas for the many-body Chern number:
\begin{align}
  C = \frac{d}{d\phi} \text{arg } \mathcal{T}(\phi;s) .
\label{Ceqn}
\end{align}

\begin{figure*}
    \includegraphics[width=1\textwidth]{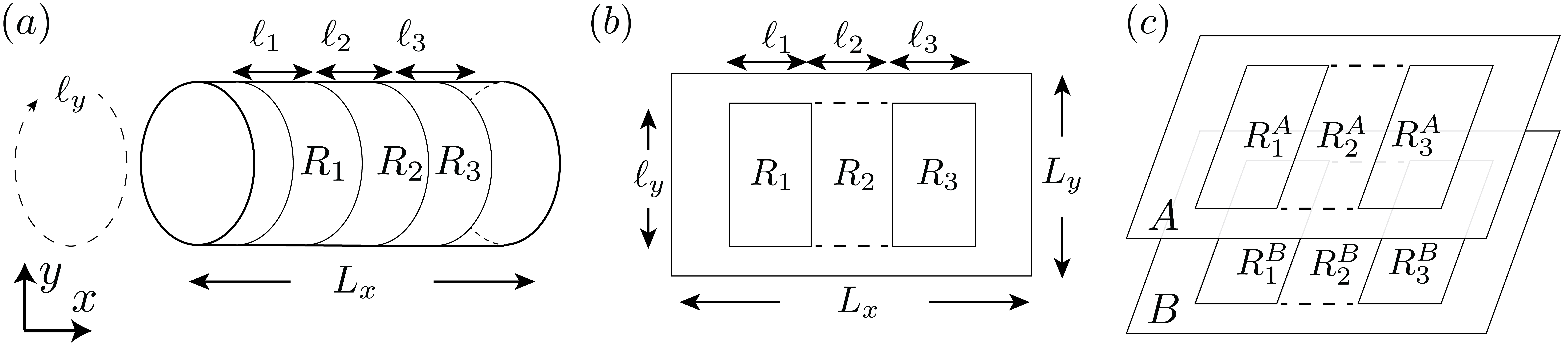}\\
    \caption{Spatial geometries corresponding to the one-layer and bilayer formulas \eqref{T1}, \eqref{T2} on a cylinder or rectangle. (a) We impose periodic boundary conditions along the $y$-direction corresponding to a cylinder in the $x-y$ space, and have the $R_i$ regions wrap all along the $y$-cycle so that $\ell_y = L_y$. (b) We impose open boundary condition along the $y$-direction corresponding to a rectangle in the $x-y$ space, and have the $R_i$ regions to be rectangular. In (a) and (b) to evaluate \eqref{T1} regions $R_1$ and $R_3$ in the bra and ket states are swapped, therefore, $\ell_3 =\ell_1$. (c) Given a copy A and B of the wave function on the same rectangular geometry, to evaluate \eqref{T2}, the swap operator is applied to regions $R_1^A$ and $R_1^B$ and to $R_3^A$ and $R_3^B$ of the two wave functions. Similarly, one can impose periodic boundary conditions on the two copies of the wave function.}
    \label{fig:spatialTopo}
\end{figure*}

Here, $\text{arg } \mathcal{T}(\phi;s) = \text{Im} \ln \mathcal{T}(\phi;s)$, and $\mathcal{T}(\phi;s)$ is defined below. Note that $\text{arg } \mathcal{T}(\phi;s)$ is linear in $\phi$ (in the thermodynamic limit), so that the slope can be calculated at any value of $\phi$. This equation is reminiscent of formulas for measuring Chern number using polarization as a function of the twist angle on a cylinder. However $\mathcal{T}(\phi;s)$ defined here depends only on a single wave function, independent of any underlying Hamiltonian.

We note that for finite-size systems, it is more robust to compute the average Chern number from the winding number,
\begin{widetext}
\begin{align}
C = \frac{1}{2\pi} \oint d\phi \frac{d}{d\phi} \text{arg } \mathcal{T}(\phi;s) ,
\end{align}
where $\phi$ winds from $0$ to $2\pi$. For $\mathcal{T}(\phi;s)$, we can consider two classes of formulas, based on using either a single copy of the state or two copies of the state:
  \begin{align}
    \label{T1}
    \mathcal{T}(\phi;s) = \langle 0| W ^\dagger_{R_1}(\phi) \mathbb{S}_{1,3} W_{R_1}(\phi) V^s_{R_1\cup R_2}|0\rangle ,
\end{align}
\begin{align}
\label{T2}
  \mathcal{T}(\phi;s) = \langle 0_A  0_B| W_{R_1^A}^\dagger(\phi) V_{ R_2^B}^{s\dagger} 
  \mathbb{S}_{1_A, 1_B}  \mathbb{S}_{3_A,3_B} W_{ R_1^A}(\phi) V_{ R_2^A}^s  |0_A 0_B\rangle .
\end{align}
\end{widetext}

Let us first consider Eq.~\eqref{T1}, which requires a single copy of the state. We have picked three subregions of the space, $R_1$, $R_2$, and $R_3$, which can be either cylindrical or rectangular, as shown in Fig.~\ref{fig:spatialTopo}(a) and Fig.~\ref{fig:spatialTopo}(b), respectively:
\begin{align}
R_i = \{(x,y) | x_{i} \leq x < x_{i+1}, y_1 \leq y \leq y_2 \}.
  \end{align}
  In the case where the regions are cylindrical as in Fig. ~\ref{fig:spatialTopo}(a), we take $y_1 = 0$ and $y_2 = L_y \sim 0$. 
We require $R_1$ and $R_3$ to have the same lengths along the $x$-direction:
 \begin{align}
  \ell_i &\equiv |x_{i+1} - x_i|, \;\; i = 1,2,3 ,
  \nonumber \\
  \ell_y &\equiv |y_2 - y_1| . 
  \end{align}
When the regions are cylindrical, $\ell_y = L_y$. 

$\mathbb{S}_{1,3}$ is a SWAP operator that swaps every particle in $R_1$ to its corresponding point in $R_3$ and vice versa. It can be written as
  \begin{align}
\mathbb{S}_{R_1,R_3} = \prod_{(x,y) \in R_1} \text{SWAP}( (x,y) , (x', y) ),
  \end{align}
  where $x' = x + (x_3 - x_1)$. Here $\text{SWAP}(\vec{r}, \vec{r}')$ moves any particle at $\vec{r}$ to $\vec{r'}$ and vice versa. Note that if we consider the Hilbert space $\mathcal{H}_R$ of the system restricted to a region $R$, we need $\mathcal{H}_{R_1}$ and $\mathcal{H}_{R_3}$ to be isomorphic so that we can define the above SWAP operator.

The operators $V_{R}$ and $W_{R}(\phi)$, which have support in region $R$, take the form
\begin{align}
  W_{R}(\phi) &= \prod_{(x,y) \in R} e^{i \hat{n}(x,y) \phi}
  \nonumber \\
   V_{R}&= \prod_{(x,y) \in R} e^{i\frac{2\pi y}{\ell_y}\hat{n}(x,y)}
\end{align}
where $\hat{n}(x,y)$ is the number density operator.

Eq.~\eqref{T2} is defined by considering two identical copies of the given state, $|0_A 0_B\rangle \equiv |0\rangle_A \otimes |0\rangle_B$, where $A$ and $B$ label the two copies as in Fig. \ref{fig:spatialTopo}(c). Here, $R^I_i$ for $I = A,B$ and $i=1,2,3$ now label two identical sets of three regions for the two copies of the system. In this case as well we can consider both cylindrical and rectangular geometries; Fig.  \ref{fig:spatialTopo}(c) shows the rectangular case. 
Eq.~\eqref{T2} is written entirely in terms of local operators, at the expense of requiring two copies of the system. This is reminiscent of the approach in Ref. \onlinecite{zhu2017} to extract fractional statistics through a product of purely on-site unitary operations by considering multiple copies of a topologically ordered state. It is also analogous to the proposal of Ref. \cite{Cirac2012order} for extracting SPT invariants using two-copies of the state. 

Eq.~\eqref{T1} and \eqref{T2} are special cases of a more general set of formulas for $\mathcal{T}(\phi;s)$:
\begin{widetext}
  \begin{align}
    \label{T1U}
    \mathcal{T}(\phi;s) = \left\{ \frac{\langle 0| \left( W_{R_1}^{a}(\phi) V_{R_1}^{s c} \right) ^\dagger \mathbb{S}_{1,3} \left( W_{R_1}^{a}(\phi) V_{R_1}^{s c}\right) W_{R_1\cup R_2}^{b}(\phi) V_{R_1\cup R_2}^{s d}|0\rangle}
  {\langle 0| \left( W_{R_1}^{a}(\phi)\right)^\dagger \mathbb{S}_{1,3} W_{R_1}^{a}(\phi) W_{R_1\cup R_2}^{b}(\phi)|0\rangle} \right\}^{1/\det U},
  \end{align}
  \begin{align}
\label{T2UR3}
    \mathcal{T}(\phi;s) = \left\{ \frac{\langle 0_A 0_B | \left[W_{R_1^A}^a(\phi) V_{R_1^A}^{sc} V^{sd}_{R_2^B} W_{R_2^B}^b(\phi)\right]^\dagger
  \mathbb{S}_{1_A, 1_B} \mathbb{S}_{3_A, 3_B}\left[W_{R_1^A}^a(\phi) V_{R_1^A}^{sc} V^{sd}_{R_2^A}  W_{R_2^A}^b(\phi)\right] |0_A 0_B\rangle}
{\langle 0_A 0_B| \left[W_{R_1^A}^a(\phi) W_{R_2^B}^b(\phi)\right]^\dagger 
    \mathbb{S}_{1_A, 1_B} \mathbb{S}_{3_A, 3_B}\left[W_{R_1^A}^a(\phi) W_{R_2^A}^b(\phi)\right] |0_A 0_B\rangle }
\right\}^{1/\det U}.
\end{align}
\end{widetext}
where $U = \left(\begin{matrix} a & b \\ c & d \end{matrix} \right)$ is a $\text{GL}(2,\mathbb{Z})$ matrix (i.e. $a,b,c,d \in \mathbb{Z}$ and $ad - bc \neq 0$). 
Note that any choice of $U \in \text{GL}(2,\mathbb{Z})$ can be used to obtain the Chern number. Eq. \eqref{T1U}-\eqref{T2U} therefore yield an infinite number of different formulae which in principle can be used to obtain the Chern number. 

If $b = 0$, the denominator in Eq. \eqref{T1U}-\eqref{T2UR3} is real, and therefore does not contribute to the Chern number. Eq. \eqref{T1}-\eqref{T2} correspond to the case where $U = \left(\begin{matrix} 1 & 0 \\ 0 & 1 \end{matrix} \right)$ is the identity matrix, and we have ignored the denominator.

We note that the numerator of Eq.~\eqref{T1U} for $U = \left(\begin{matrix} 0 & 1 \\ -1 & 1 \end{matrix} \right)$ and $s = 1$ was proposed in  Ref.~\onlinecite{Shiozaki2018many} for extracting the many-body Chern number for integer Chern insulators from the ground state defined on a cylinder. However, we find that the denominator is also crucial to obtain correct results. 

We find that we can also remove the swap in region $R_3$ above, to instead use the formula:
\begin{widetext}
  \begin{align}
\label{T2U}
    \mathcal{T}(\phi;s) = \left\{ \frac{\langle 0_A 0_B | \left[W_{R_1^A}^a(\phi) V_{R_1^A}^{sc} V^{sd}_{R_2^B} W_{R_2^B}^b(\phi)\right]^\dagger
  \mathbb{S}_{1_A, 1_B} \left[W_{R_1^A}^a(\phi) V_{R_1^A}^{sc} V^{sd}_{R_2^A}  W_{R_2^A}^b(\phi)\right] |0_A 0_B\rangle}
{\langle 0_A 0_B| \left[W_{R_1^A}^a(\phi) W_{R_2^B}^b(\phi)\right]^\dagger 
    \mathbb{S}_{1_A, 1_B} \left[W_{R_1^A}^a(\phi) W_{R_2^A}^b(\phi)\right] |0_A 0_B\rangle }
\right\}^{1/\det U}.
\end{align}
\end{widetext}
However we find from our numerical simulations that Eq. \eqref{T2U} is not as robust as Eq. \eqref{T2UR3} for finite size systems. 

In the case where the regions $R_i$ are cylindrical, we have an analytical understanding of these formulas in terms of topological quantum field theory (TQFT), which is backed up by extensive numerical simulations. However in the case where the regions $R_i$ are rectangular, we do not have an analytical understanding in terms of TQFT. Nevertheless, our numerical results indicate that the formulas continue to reproduce the Chern number.

This paper is organized as follows. In Section \ref{sec2}, we review the conventional definition of the many-body Chern number using wave functions with twisted boundary conditions. We then review how the Chern number and integer $s$ can be extracted from the many-body polarization operator. In reviewing the generalization of these concepts to FQH systems, we also present some results which to the best of our knowledge have not been addressed before. In Section \ref{secTQFT}, we explain our TQFT approach for the derivation of the many-body Chern number when the regions $R_i$ are cylindrical. After reviewing the $U(1)$ Chern-Simons response theory, we reinterpret the many-body polarization of the Chern number from a topological field theory viewpoint. Next, we demonstrate how the expectation value of the SWAP operator for a wave function defined on a torus can be interpreted in the TQFT in terms of the path integral on a space-time 3-torus. This allows the introduction of two non-contractible cycles that are both orthogonal to the real time direction. The introduction of symmetry defect operators $V$ and $W$ using just the density operator can then be used to introduce the appropriate non-trivial background gauge field configurations that allow extraction of the Chern number. 
In Section \ref{sec:numerical}, we provide extensive numerical evidence
establishing that the Chern number can indeed be extracted from a single wave function. Our numerical studies include a number of bosonic and fermionic FQH states, with Abelian and non-Abelian topological orders, and for different system sizes. We also demonstrate that our formulas can detect the occurrence of topological phase transitions. Next, we study the effect of changing the support of the symmetry defect operators $W_{R}(\phi)$ and $V_R$ for the case where the regions $R_i$ are cylindrical. We end this section with a numerical study of the system-size dependence of the magnitude of our SWAP formulas. 
We conclude with a discussion in section \ref{secDiscussion} and give additional details of our proofs in four appendices.

\section{Chern number from Berry phase}\label{sec2}

In this section, we discuss the conventional definition of the many-body Chern number and its relation to polarization in one dimension. While some of this section is review, the discussion in the context of the FQH states is more general than other treatments and contains some results that are not, to our knowledge, discussed elsewhere in the literature. 

\subsection{Chern number from twisted boundary conditions on torus}

Consider a many-body system defined on a torus, with spatial periodicity $L_x$ and $L_y$ in the $x$ and $y$ directions, respectively. We consider the ground state of the system in the presence of flux $\theta_x$ and $\theta_y$ (in units where flux $2\pi$ is equal to the flux quantum) through the two non-contractible cycles of the torus. Equivalently, upon performing a singular gauge transformation, we consider the ground state $|\Psi(\theta_x, \theta_y)\rangle$ with twisted boundary conditions:
\begin{align}
    \hat{t}_{i}^{(j)}(L_i)|\Psi(\theta_x, \theta_y)\rangle = e^{i\theta_i}|\Psi(\theta_x, \theta_y)\rangle, 
\end{align}
where $i={x,y}$ and $\hat{t}_{i}^{(j)}$ denotes the single-particle magnetic translation operator for the $j$th particle with a displacement $L_i$ along the $i$th direction. 

\subsubsection{Integer quantum Hall states}

When there is an energy gap above a non-degenerate ground state, gauge invariance requires that adiabatically varying each flux from $0$ to $2\pi$ takes the state back to itself. The Berry connection is defined as
\begin{align}
\mathcal{A}_i(\theta_x, \theta_y) = -i \langle\Psi| \frac{\partial}{\partial \theta_i}|\Psi\rangle . 
\end{align}
The Chern number is then defined to be
\begin{align}
  \label{ChernNumberTorus1}
C = \frac{1}{2\pi} \int d^2 \theta \mathcal{F}(\theta_x, \theta_y) ,
\end{align}
where the integration is over $0 \leq \theta_i \leq 2\pi$, and the Berry curvature is
\begin{align}
\mathcal{F}(\theta_x, \theta_y) = \partial_{\theta_x}\mathcal{A}_y - \partial_{\theta_y}\mathcal{A}_x . 
\end{align}
The Chern number defined above is quantized to be an integer. Abstractly, the integral in Eq.~\eqref{ChernNumberTorus1} evaluates the first Chern class of a principal bundle with a $U(1)$ structure group over a two-torus $T^2$, which must be an integer. 

It can be shown that the Hall conductivity is given by
\begin{align}
\sigma_H = \frac{e^2}{h} 2\pi  \mathcal{F}(\theta_x, \theta_y) . 
\end{align}
The fact that $\mathcal{F}(\theta_x, \theta_y) $ is itself quantized up to exponentially small corrections in $\xi/L_i$, where $\xi$ is the correlation length, without the need to average over the space of boundary conditions, follows from the spectral gap of the Hamiltonian and exponentially decaying correlations, as rigorously proven  in Ref. \onlinecite{hastings2015}, and numerically verified in Ref. \onlinecite{Kudo2019Many}.

For non-interacting, translationally invariant systems, the above definition of the Chern number reduces to the TKNN invariant \cite{Thouless1982Quantized}, which assigns an  invariant to each band by averaging the Berry curvature of single-particle Bloch wave functions in momentum space.

\subsubsection{Fractional quantum Hall states}

Topologically ordered states of matter, such as fractional quantum Hall states, possess topologically degenerate ground states on a torus \cite{Wen1990Ground}, and a fractionally quantized Hall conductivity
\begin{align}
\sigma_H = \frac{p}{q} \frac{e^2}{h},
\end{align}
for $p$ and $q$ coprime. Thus, the above discussion must be modified \cite{hatsugai2005characterization}.  Instead of a single state, we consider the full multiplet of $M$ topologically degenerate ground states on a torus, and we define an integer Chern number $\tilde{C}$ as follows. The non-Abelian Berry connection is given
\begin{align}
\mathcal{A}_{ab} = -i \langle a ; (\theta_x,\theta_y) | \partial_{\theta} | b; (\theta_x,\theta_y) \rangle ,
\end{align}
where $a,b = 0,\cdots,M-1$ label the $M$ states on a torus in an arbitrary choice of basis. 
This defines a gauge field for a $U(M)$ bundle on the torus with field strength $\mathcal{F} = d \mathcal{A} + \mathcal{A} \wedge \mathcal{A}$. The trace gives an integer Chern number
\begin{align}
  \tilde{C} = \frac{1}{2\pi} \int d^2\theta_x \text{Tr } \mathcal{F}.
\end{align}

The Hall conductivity corresponds to the Berry curvature averaged over the $M$ degenerate states. Including the average over the space of boundary conditions then gives
\begin{align}
  \sigma_H &= \frac{\tilde{C}}{M} \frac{e^2}{h}  = \frac{p}{q} \frac{e^2}{h}. 
\end{align}
As in the IQH case, one uses the fact that the Hall conductivity is independent of boundary conditions due to the exponentially decaying correlations of the system, and thus one can average over the space of boundary conditions to relate the Hall conductivity directly to the Chern number \cite{hastings2015}.

Note that the above calculation requires  the entire multiplet of the ground states on a torus, and also how they evolve as the flux $\theta_x$ and $\theta_y$ change from $0$ to $2\pi$. A natural question is thus whether the Hall conductivity can be computed with less information. 

In fact, one does not need to use all of the $M$ states. In particular, the structure of FQH states on a torus is such that the degenerate $M$ ground states break up into $k$ sectors. Within each sector  $i$, there are degenerate states $|\alpha,i \rangle$, for $i = 1,\cdots, k$, and $\alpha = 0,\cdots, m_i - 1$. Here $m_i$ is the number of ground states within the $i$th sector. 
  
The key defining property of these sectors is that the states in a given sector are related to each other under the operation of flux insertion. Concretely, if we define operators $\hat{\Phi}_x$ and $\hat{\Phi}_y$ that adiabatically insert $2\pi$ flux through the $x$ and $y$ cycles of the torus respectively (followed by a large gauge transformation to remove the flux), we can pick a basis of states such that
\begin{align}
  \label{fluxIns}
  \hat{\Phi}_x | \alpha,i \rangle &= | (\alpha+1) \text{ mod } m_i,i \rangle
  \nonumber \\
  \hat{\Phi}_y | \alpha,i \rangle &=e^{i\gamma_i} e^{i 2 \pi \alpha p/q }|\alpha,i \rangle ,
\end{align}
for some $\alpha$-independent phase $\gamma_i$.
Note that the algebra of the flux insertion operators in Eq. (\ref{fluxIns}) requires that $m_i$ is an integer multiple of $q$:
\begin{align}
m_i= r_i q, 
\end{align}
where $r_i$ is a positive integer. 

As an example, consider a bilayer state consisting of a bosonic $1/2$ Laughlin state in each layer. Such a FQH state has $4$ topologically degenerate ground states on a torus, which can be labeled $|ab\rangle$, for $a,b = 0,1$. Under flux insertion, $|a,b\rangle \rightarrow |(a+1) \text{ mod } 2, (b+1) \text{ mod } 2\rangle$, so that in this example $M = 4$, $k =2$, and $m_1 = m_2 = 2$. Alternatively, the bosonic Moore-Read Pfaffian state at $\nu = 1$ has $M = 3$, $k = 2$, $m_1 = 2$, $m_2 = 1$. 

Thus we can define a Chern number for each sector by defining the states as a function of the twist angle, $|\alpha,i;(\theta_x,\theta_y)\rangle$, and considering the Berry connection and curvature:
\begin{align}
  \mathcal{A}_{\alpha\beta}^{(j)} &= -i \,\langle \alpha, j; (\theta_x,\theta_y)| \partial_{\theta} |\beta, j; (\theta_x, \theta_y)\rangle
  \nonumber \\
  \mathcal{F}^{(j)} &= d \mathcal{A}^{(j)} + \mathcal{A}^{(j)} \wedge \mathcal{A}^{(j)}
\end{align}
We then have
\begin{align}
  C_i &=  \frac{1}{2\pi} \int_0^{2\pi} d \theta_x \int_0^{2\pi} d \theta_y \text{Tr } \mathcal{F}^{(i)} ,
\end{align}
where the trace is taken within the $i$th sector. Since the Hall conductivity is a local observable \cite{hastings2015} and the different degenerate ground states are indistinguishable by local operators, we thus expect averaging over all of the degenerate topological ground states should give the same result as just averaging over a particular sector. Therefore, 
\begin{align}
\frac{C_i}{m_i} = \frac{\tilde{C}}{M} ,
\end{align}
which means that $C_i/m_i$ is independent of $i$. 

Instead of following $m_i$ states as both twist angles $\theta_x,\theta_y$ are varied from $0$ to $2\pi$,  an alternative way of obtaining $C_i$ is to follow a single state, as a single twist angle $\theta_x$ is varied from $0$ to $2\pi m_i$. Adiabatic insertion of flux only returns a state in the $i$th sector back to itself after $m_i$ units of flux have been inserted. Therefore, we can consider the states as a function of the twist angle through the two holes, $| \alpha,i; (\theta_x,\theta_y) \rangle$, and we can pick a basis such that
\begin{align}
|\alpha,i; (\theta_x + 2\pi, \theta_y) \rangle = | (\alpha+1) \text{ mod } m_i,i; (\theta_x, \theta_y) \rangle . 
  \end{align}
  Thus we can define
  \begin{align}
C_i = \frac{1}{2\pi} \int_{0}^{2m_i \pi}d\theta_x \int_{0}^{2\pi}d\theta_y \mathcal{F}_{\alpha,i}(\theta_x, \theta_y) ,
  \end{align}
  where $\mathcal{F}_{\alpha,i}(\theta_x, \theta_y)$ is the $U(1)$ Berry curvature in the space of fluxes $(\theta_x,\theta_y)$ for the given state $|\alpha,i; (\theta_x,\theta_y) \rangle$. 
 Since flux insertion from $0$ to $2\pi m_i $ cycles the state through every choice of $\alpha$, it is clear that $C_i$ is independent of the choice of $\alpha$.

At this stage it is useful to note that in general,
\begin{align}
  s = \text{max}_i \; m_i = \text{lcm}( \{m_i\}) = rq,
\end{align}
for some integer $r$, which we prove in Appendix \ref{zlcm}. Here lcm refers to the least common multiple. Thus we can define
\begin{align}
  \label{Cdef}
C \equiv s \frac{C_i}{m_i} = s p / q = r p,
\end{align}
in terms of which the Hall conductivity is
\begin{align}
\sigma_H = \frac{C}{s} \frac{e^2}{h} .
\end{align}

The usefulness of $C$ can be seen by observing that the calculations sketched above require either (a) knowledge of how all $M$ ground states evolve as the twist angles $(\theta_x,\theta_y)$ are varied, or (b) knowledge of a specific basis that satisfies (\ref{fluxIns}) for a given sector. Suppose instead that we have access to a single state on a torus $|\Psi(\theta_x,\theta_y)\rangle$ as a function of the twist angles, which corresponds to an arbitrary superposition:
\begin{align}
|\Psi(\theta_x,\theta_y)\rangle = \sum_{i=1}^k \sum_{\alpha = 0}^{m_i-1} \Psi_{\alpha,i} |\alpha,i; (\theta_x,\theta_y) \rangle .
\end{align}
This state is guaranteed to be periodic for $(\theta_x,\theta_y) \rightarrow (\theta_x + 2\pi s, \theta_y)$ and $(\theta_x,\theta_y) \rightarrow (\theta_x, \theta_y + 2\pi s)$. Thus we can define an integer Chern number by defining a $U(1)$ Berry connection and curvature:
\begin{align}
  \label{Cpsi}
   \mathcal{A}^\psi &= -i \langle \Psi(\theta_x,\theta_y) | \partial_{\theta} |\Psi(\theta_x,\theta_y)\rangle
           \nonumber \\
  \mathcal{F}^\psi &= \partial_{\theta_x} A_y^\psi - \partial_{\theta_y} A_x^\psi
           \nonumber \\
  C^\psi &= \frac{1}{2\pi s} \int_0^{2\pi s} d\theta_x \int_0^{2\pi s} d\theta_y \mathcal{F}^\psi .
\end{align}

Again, because the Hall conductivity is a local observable \cite{hastings2015}, we expect to obtain the same result regardless of which state on the torus we pick to compute the Hall conductivity. Consequently, we expect
$C^\psi/s = C_i/m_i$, or equivalently
\begin{align}
  \label{CpsiCeq}
C^\psi = C,
  \end{align}
  where $C$ is defined in Eq. \ref{Cdef}.

In order to numerically calculate or physically measure the above quantities, one needs, at the very least, to have access to a continuous family of wave functions in the neighborhood of any given point $(\theta_x, \theta_y)$, which, depending on the resources involved, may be experimentally or computationally intensive. The main result of this paper is to show how to compute $C$ given $s$ and a single wave function on a disk. In order to understand these results, it is important to review the known method to extract the Chern number from the polarization of a state as a function of a single parameter $\theta_x$.

\subsection{Chern number from many-body polarization}
\label{polarizationSec}

In this section, we briefly review the Resta formula \cite{KingSmith1993Theory, Resta1994Macroscopic,Resta1998Quantum} on the relation between the many-body polarization operator and the Chern number and generalize it to the case where the ground state subspace is degenerate.

\subsubsection{IQH states}

We first consider an IQH state on a torus. We take the coordinates to be $(x,y) \in [0, L_x] \times [0,L_y]$, where the $x$ and $y$ directions are compactified. Let $|\Psi(\theta_x)\rangle$ be the ground state wave function for an IQH state in the presence of a flux through the $x$ direction
\begin{align}
\oint dx A_x = \theta_x .
\end{align}
Without loss of generality, we take the flux in the $y$ direction to be zero, $\oint dy A_y = 0$. We note that for the following argument, one can also consider a cylinder instead of a torus, although the torus is slightly more convenient in discussing the FQH case below. Following Resta, we define the exponentiated polarization operator
 \begin{align}
R_y = \prod_{x,y} e^{i \frac{2 \pi y}{L_y} \hat{n}(x,y)}, \label{eq:resta_op}
\end{align}
where the product is taken over the whole system. We then compute
\begin{align}
\mathcal{T}(\theta_x) = \frac{\langle \Psi(\theta_x) | R_y | \Psi(\theta_x) \rangle}{\langle \Psi(\theta_x) |  \Psi(\theta_x) \rangle}.  
\end{align}

Adiabatically changing $\theta_x$ is equivalent to applying an electric field $E_x$, which induces a current in the $\hat{y}$ direction due to the Hall conductivity, which corresponds to a changing polarization along the $\hat{y}$ direction. The Chern number therefore can be obtained as 
\begin{align}
C =  \frac{d}{d\theta_x} \text{arg } \mathcal{T}(\theta_x). 
\end{align}

A more robust quantity for finite size systems is the winding number
\begin{align}
C = \frac{1}{2\pi} \oint d \theta_x \frac{d}{d\theta_x} \text{arg } \mathcal{T}(\theta_x).
\end{align}

\subsubsection{FQH  states}

The discussion in the case of the FQH states is again complicated by the existence of $M$ independent, topologically distinct states on a torus. 

Let us consider a state $|\Psi (\theta_x) \rangle$ on the torus as before, which now can correspond to any possible superposition of the different topological sectors:
\begin{align}
|\Psi(\theta_x)\rangle = \sum_\alpha \sum_i \Psi_{\alpha, i}| \alpha,i; \theta_x \rangle , 
\end{align}
where the flux corresponds to the holonomy of $A$ along the $x$ direction.

The complication now is that in general $R_y$ has a non-trivial action in the degenerate ground state subspace, and therefore $R_y |\Psi(\theta_x)\rangle$ may be orthogonal to $|\Psi(\theta_x)\rangle $ \cite{Aligia1999Quantum, Nakamura2002Lattice}.
Neverthless, in the thermodynamic limit,
\begin{align}
  \label{RyPsi}
 R_y^s |\Psi(\theta_x)\rangle \propto |\Psi(\theta_x) \rangle,
\end{align}
where $s$ was defined in the previous section. We provide a derivation of this in Appendix \ref{RysPsi}.

Consequently, we consider
\begin{align}
  \label{TaPol}
\mathcal{T}(\theta_x; s) = \frac{\langle \Psi(\theta_x) | R_y^s |\Psi(\theta_x) \rangle}{\langle \Psi(\theta_x) | \Psi(\theta_x) \rangle} , 
\end{align}
and then define
\begin{align}
C = \frac{d}{d\theta_x} \text{arg } \mathcal{T}(\theta_x; s) . 
  \end{align}

The Hall conductivity then corresponds to
\begin{align}
\sigma_{H} = \frac{C}{s} \frac{e^2}{h} . 
\end{align}

\subsection{Extracting $s$ from ground state wave functions}

In general our formulas require knowledge of an additional topological invariant, $s$, in order to obtain the many-body Chern number. Here we briefly discuss how $s$ can be obtained given the degenerate set of ground state wave functions on a
torus, without knowledge of the Hamiltonian and without considering the wave functions in the space of twist angles.

Recall that $s$ is the minimal integer that satisfies, in the thermodynamic limit,
\begin{align}
R_y^s | \Psi \rangle = e^{i \lambda} | \Psi \rangle,
\end{align}
for any state on the torus, where $\lambda$ is a global phase factor. It follows that given the degenerate set of ground state wave functions on a torus, one can systematically search for $s$ by applying increasing powers of $R_y$ to the ground states.

We note that one could replace the torus with the cylinder, in which case we would require a representative ground state from each of the $M$ topological sectors on the cylinder.

We have verified the above statement numerically in a number of examples, including the non-Abelian bosonic Moore-Read Pfaffian state at $\nu  =1$, for which $s = 2$ (see Appendix \ref{RysPsi}).

\section{Topological quantum field theory approach}\label{secTQFT}

We now wish to derive Eqs.~\eqref{T1},\eqref{T2},\eqref{T1U},\eqref{T2UR3},\eqref{T2U} using insights from topological quantum field theory (TQFT), which describes the low energy, long wavelength universal properties of topological phases of matter \cite{atiyah1988topological,Witten1989,wen04,nayak2008}. The results of this section will lead to a TQFT-based derivation of Eqs.~\eqref{T1},\eqref{T2},\eqref{T1U},\eqref{T2UR3},\eqref{T2U} for the case where the regions $R_i$ are cylindrical. 

In Sec. \ref{TQFTrev} we first briefly review some of the main tools from TQFT that we use. In Sec.~\ref{secCSreponse}, we interpret and generalize the polarization formula for the many-body Chern number through the lens of TQFT and the Chern-Simons response theory and discuss a number of related issues. We show that calculating the Chern number via the many-body polarization can be understood in terms of the path integral of the TQFT on $S^1 \times S^2$ decorated with two symmetry defects corresponding to the non-trivial background gauge field configurations. 

In Sec.~\ref{surgerySec}, we show how to construct the TQFT path integral on topologically non-trivial space-time manifolds with the aid of SWAP operations. For example, for (1+1)D TQFTs, we show how the path integral on a space-time torus $T^2$ can be obtained by starting with the wave function on a circle $S^1$. In Sec.~\ref{secSymmetryDefect}, we explicitly present the form of the inserted  symmetry  defect operators that effectively induce the appropriate non-trivial background gauge field configurations. By combining this with the results of Sec.~\ref{secCSreponse} and \ref{surgerySec}, we then arrive at our formulas for the case where the regions $R_i$ are cylindrical. 

Note that in this section we pick units $e = \hbar = 1$, such that the Hall conductivity is $\sigma_H = \frac{1}{2\pi} \frac{p}{q}$. 

\subsection{Review of TQFT essentials}
\label{TQFTrev}

\subsubsection{Path integrals and surgery}

Given any closed space-time manifold $M$, the TQFT defines a topologically invariant path integral $\mathcal{Z}(M)$. In the presence of a $U(1)$ symmetry, the theory can also be coupled to a background $U(1)$ gauge field $A$, and the TQFT thus defines a topologically invariant path integral $\mathcal{Z}(M; A)$. A simple example is the TQFT obtained by quantizing a CS theory with a dynamical (emergent) $U(1)$ gauge field $a$, for which the path integral is formally written as
\begin{align}
  \label{CSgaugeT}
\mathcal{Z}(M; A) = \int \mathcal{D} a e^{i \int_M ( -\frac{m}{4\pi} a d a + \frac{1}{2\pi} A da )} ,
\end{align}
where in a local coordinate patch $a d a \equiv \epsilon^{\mu\nu\lambda} a_\mu \partial_\nu a_\lambda d^3 x$ \footnote{To be precise, when non-trivial $U(1)$ bundles are involved, the $U(1)$ CS theory is most properly defined by viewing $M$ as the boundary of a $4$-manifold $W$, extending the $U(1)$ gauge fields onto $W$, and defining the action in terms of topological $\theta$ terms using only the field strengths \cite{Dijkgraaf1990}.}. The above TQFT describes the $1/m$ Laughlin state. 

On a manifold $M$ with boundary, the TQFT defines a state on the boundary $\partial M$, $|\Psi_{\partial M}\rangle$, which can also be thought of as the value of the path integral on $M$. In the above example, we can formally obtain a wave function
\begin{align}
\Psi_{\partial M}( \tilde{a}, A) = \int_{a|_{\partial M} = \tilde{a}} \mathcal{D} a e^{i \int_M ( -\frac{m}{4\pi} a d a + \frac{1}{2\pi} A da )} 
  \end{align}
in terms of the path integral with fixed boundary conditions on $\partial M$. For example, the path integral on the solid torus, $\mathcal{Z}(S^1 \times D^2)$, determines a state on the torus, a state on a torus,  $|\Psi_{T^2}\rangle$, as shown in Fig. \ref{fig:symDefects}(a). 

The TQFT also possesses a gluing formula. Suppose that a closed manifold $M$ is obtained by gluing together two manifolds $M_1$ and $M_2$ along their boundary according to the homeomorphism $f: \partial M_1 \rightarrow \partial M_2$, such that $M = M_1 \cup_f M_2$. Then, the path integral on $\mathcal{Z}(M)$ corresponds to
\begin{align}
\mathcal{Z}(M) = \langle \Psi_{\partial M_1} | \hat{\Lambda}_f | \Psi_{\partial M_2} \rangle,
\end{align}
where $\hat{\Lambda}_f$ is the representation of the gluing map $f$ on the quantum Hilbert space. For example, using the identity map on the solid torus restricted to the boundary, we have
\begin{align}
  \mathcal{Z}(S^2 \times S^1) = \langle \Psi_{T^2}| \Psi_{T^2} \rangle  \label{eq:Z(S2,S1)} 
\end{align}
as illustrated in Fig. \ref{fig:symDefects}(a).

\subsubsection{Chern-Simons response theory}

The low frequency, long wavelength electromagnetic response of the system at low temperatures can be encoded in an effective action for the electromagnetic gauge field, such that the path integral of the TQFT $\mathcal{Z}(M; A)$ as a function of the background gauge field $A$ is given by
\begin{align}
\label{responseTheory}
\mathcal{Z}(M; A) = \mathcal{Z}(M; 0) e^{i \frac{p}{q} S_{CS}[A]} ,
\end{align}
where the Chern-Simons response action is given by
\begin{align}
\label{CSresponseAction}
S_{CS}[A] = \frac{1}{4\pi} \int_M A d A,
\end{align}
where recall that in a local coordinate patch $A d A \equiv \epsilon^{\mu\nu\lambda} A_\mu \partial_\nu A_\lambda d^3x$. 

Note that in the above discussion, $\mathcal{Z}$ is the path integral of the TQFT which describes the long wavelength universal properties of the system. In the path integral of the full microscopic theory, $S_{CS}$ appears as the leading order term in an expansion in gauge-invariant combinations of $A$ and its gradients. The subleading terms consist of increasing powers and derivatives of the field strength $F_{\mu\nu} = \partial_\mu A_\nu - \partial_\nu A_\mu$, such as the Maxwell term.

The CS response theory thus tells us how to obtain $\mathcal{Z}(M; A)$ in terms of $\mathcal{Z}(M; 0)$. The full TQFT thus includes additional prescriptions for how to compute $\mathcal{Z}(M; 0)$, for example from quantizing CS gauge theories with appropriate dynamical (emergent) gauge fields as in Eq. \eqref{CSgaugeT}.

Note that Eq. \eqref{responseTheory} is only well-defined for certain classes of $U(1)$ bundles (this is related to the necessity of including $s$ in Eq. \eqref{numeratorGauge} below). Furthermore, Eq. \eqref{CSresponseAction} is strictly speaking well-defined only for trivial $U(1)$ bundles; for non-trivial $U(1)$ bundles, a proper definition of the CS term often requires defining the theory as the boundary of a (3+1)D theory \cite{Dijkgraaf1990}. Properly evaluating the CS response term therefore requires some care, as we will see through our calculations.

\subsubsection{Symmetry defects}\label{sec:symmetryDefects}

In order to extract the Chern number, we need to not only create a non-trivial topology for the space-time manifold, but we also need to obtain the path integral of the TQFT for certain non-trivial $U(1)$ bundles; that is, for certain families of background $U(1)$ gauge field configurations $A$. 

Such a path integral can be obtained by inserting symmetry defect operators into the correlation functions in the TQFT. The symmetry defect operators are codimension-1 operators in space-time that effectively impart phase jumps in the local trivializations that define the $U(1)$ bundle. Equivalently, they effectively change the gauge field configuration in the following way. Suppose that a symmetry defect operator induces a phase jump by $\phi$ as it is crossed. Let $\hat{u}$ be the unit vector normal to the region of support of the operator. Then, the gauge field configuration is effectively changed by
\begin{align}
\delta A = \phi \delta(u) \hat{u} ,
\end{align}
where $u$ here is the coordinate along the $\hat{u}$ direction and the symmetry defect is located at $u = 0$.

Symmetry defects are essential for understanding how to couple TQFTs to background $G$ bundles for any symmetry group $G$, and have recently played an important role in understanding how to fully characterize symmetry in topological phases of matter. Ref. \onlinecite{Barkeshli2019Symmetry}, for example, developed an algebraic theory of symmetry defects for (2+1)D topological orders. 

\subsection{Chern number from TQFT path integrals}
\label{secCSreponse}

\subsubsection{Chern number from polarization and $M  = S^2 \times S^1$}
\label{CSpol}

Here we reinterpret the calculation of the Chern number in terms of the many-body polarization in Section \ref{polarizationSec} from the point of view of the path integral of the TQFT.

As discussed above, the state $|\Psi_{T^2}(\theta_x)\rangle$ on a torus in the TQFT corresponds to the path integral on the corresponding solid torus, $\mathcal{Z}(S^1 \times D^2; A)$. Note that the precise state $|\Psi_{T^2}\rangle$ that is obtained on the torus is determined in the path integral calculation by inserting the appropriate quasiparticle Wilson loop (or superposition of Wilson loops) along the topologically non-trivial cycle of $S^1 \times D^2$.

\begin{figure}[h]
    \centering 
    \includegraphics[width=.4\textwidth]{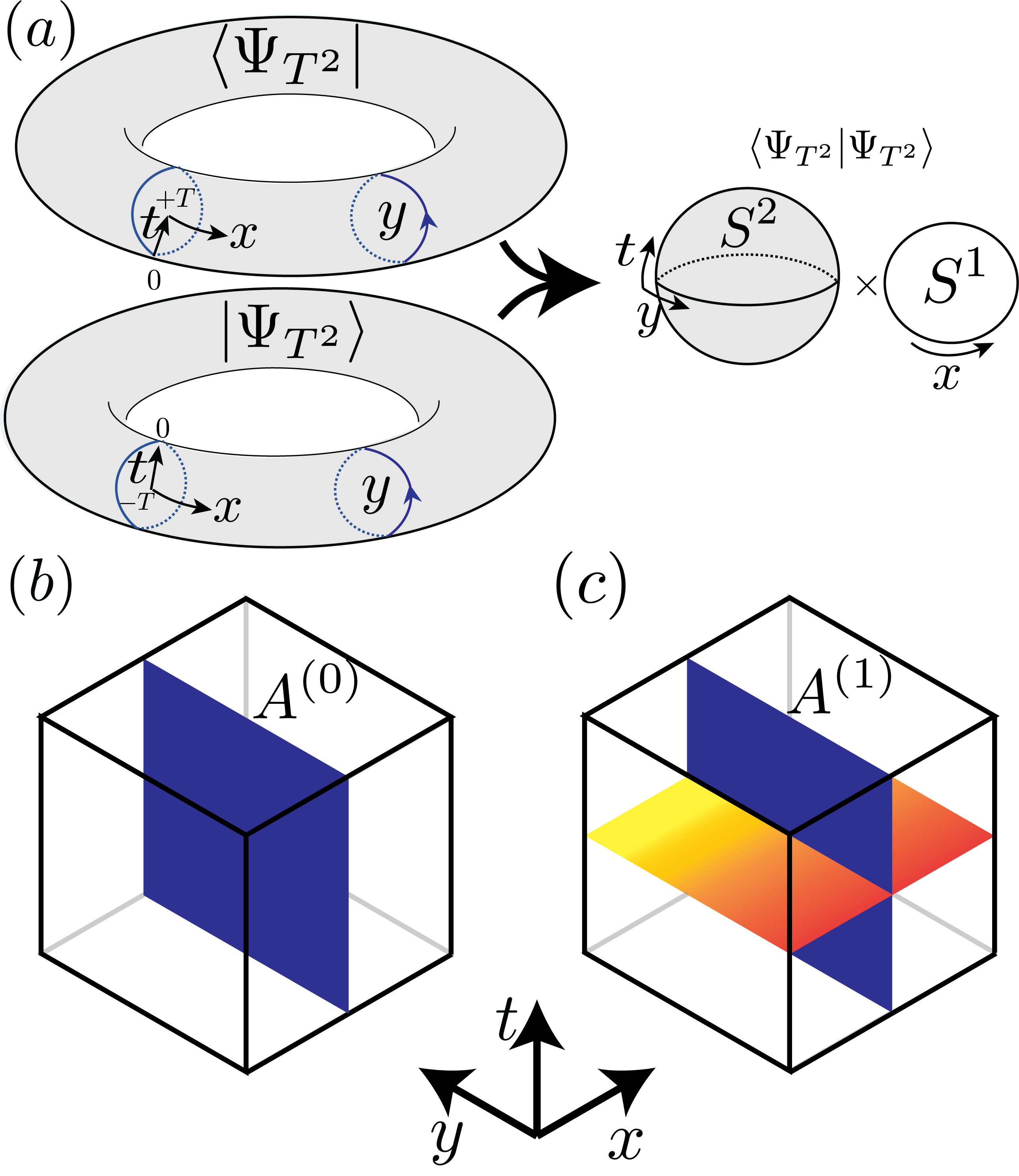}\\
    \caption{(a) The wave functions of $
    \langle \Psi_{T^2}(\theta_x)|$ and $|\Psi_{T^2}(\theta_x)\rangle$,  on a torus in the TQFT corresponds to the path integral on the solid torus, $\mathcal{Z}(S^1 \times D^2; A)$. The inner product $\langle \Psi_{T^2}(\theta_x)|\Psi_{T^2}(\theta_x)\rangle$ corresponds to evaluating the path integral on $S^1 \times S^2$.
    (b) The space-time manifold is decorated with two planar symmetry defects along the $x$ direction, described by $A_x = \theta_x \delta(x)$. The magnetic flux $\theta_x$ creates a symmetry defect described by $A_x$. (c) The electric field is generated with an additional symmetry defect along  the $x$ direction as $A_t = 2 \pi s y/L_y \delta(t)$. The coloring along $A_t$ symmetry defect illustrates its non-zero gradient. The crossing of the two symmetry defects generates a finite result for the Chern-Simons action.}
    \label{fig:symDefects}
\end{figure}

As shown in Fig.~\ref{fig:symDefects}(a), let us fix coordinates on the $S^1 \times D^2$ as $(x,y,t)$, where $x \in [-L_x/2,L_x/2]$ lies along the first $S^1$ and $(y,t)$ label the coordinates on the $D^2$. $t$ and $y$ are chosen to lie along the radial and angular directions on the $D^2$, respectively.

The state $|\Psi_{T^2}(\theta_x)\rangle$ is defined for a system with non-trivial flux, $\oint_x A_x =\theta_x$. Without loss of generality we can consider this to arise from a gauge field configuration $A_x = \theta_x \delta(x)$, $A_y = 0$. Thus the state $|\Psi_{T^2}(\theta_x)\rangle$ corresponds to evaluating the path integral on $\mathcal{Z}(S^1 \times D^2; A)$ with the gauge field configuration $(A_t,A_x,A_y) = (0,\theta_x \delta(x), 0)$, as shown in Fig.~\ref{fig:symDefects}(b).

The inner product $\langle \Psi_{T^2}(\theta_x) | \Psi_{T^2}(\theta_x) \rangle$ therefore corresponds to evaluating the path integral on $(S^1 \times D^2) \cup (S^1 \times D^2) = S^1 \times S^2$, where the two factors of $S^1 \times D^2$ are glued to each other by the trivial boundary map.

We therefore conclude that in the TQFT,
\begin{align}
  \langle \Psi_{T^2}(\theta_x) | \Psi_{T^2}(\theta_x) \rangle = \mathcal{Z}(S^2 \times S^1; A^{(0)}),
\end{align}
with the background gauge field configuration
\begin{align}
  \label{denominatorGauge}
A^{(0)} = (0, \theta_x \delta(x), 0) . 
\end{align}
Here we have picked coordinates as follows. The $x$ direction corresponds to the $S^1$, as above. The $S^2$ corresponds to two disks glued together; $y$ is taken to lie along the angular direction on each disk, while $t \in [-T,0]$ lies along the radial direction of the first disk, and $t \in [0,T]$ lies along the negative of the radial direction of the second disk. 

Since only the $x$ component of $A^{(0)}$ is non-zero, we have 
\begin{align}
S_{CS}[A^{(0)}] = 0. 
  \end{align}
Therefore, using Eq. \eqref{responseTheory}, we conclude
  \begin{align}
\mathcal{Z}(S^2 \times S^1; A^{(0)}) = \mathcal{Z}(S^2 \times S^1; 0) .
\end{align}

Let us now consider the numerator of the expectation value $\mathcal{T}(\theta_x;s)$ in Eq.~\eqref{TaPol}, and interpret the polarization operator Eq. \eqref{eq:resta_op}, as an additional symmetry defect operator in Chern-Simons theory. Specifically, the numerator corresponds to the path integral with an inserted operator
\begin{align}
e^{i \frac{2 \pi s}{L_y} \int dy dx  y \hat{n}(x,y)}= e^{i \int_M j_t A_t},
  \end{align}
where we identified it as a source term with $j_t \equiv \hat{n}$ as the microscopic density operator, and $A_t = \frac{2 \pi s y }{L_y} \delta(t)$ as the $t$-component of the gauge field. We have chosen the $t$ coordinate so that the insertion occurs at $t = 0$. In other words, we can interpret the numerator of $\mathcal{T}(\theta_x; s)$ in Eq. \eqref{TaPol} as computing the path integral of the full many-body system with the gauge field configuration
\begin{align}
  \label{numeratorGauge}
(A^{(1)}_t, A^{(1)}_x, A^{(1)}_y) =\left( \frac{2 \pi y s}{L_y} \delta(t),  \theta_x \delta(x), 0 \right), 
\end{align}
as shown in Fig.~\ref{fig:symDefects}(c). Therefore, the inner product in the numerator of Eq. \eqref{TaPol} can be written in the TQFT as
\begin{align}
  \langle \Psi_{T^2}(\theta_x) | e^{i\int_{T^2} A_t j_t} | \Psi_{T^2}(\theta_x) \rangle = \mathcal{Z}(S^2 \times S^1; A^{(1)}).
\end{align}
Note that for this expectation value not to vanish, the factor of $s$ is required in Eq. \eqref{numeratorGauge}. While we argued this in Sec. \ref{polarizationSec}, it is also possible to derive it in the TQFT description, for example for the CS gauge theory of Eq. \eqref{CSgaugeT}, although we will not do so here. 

We can equivalently interpret the path integral in the presence of this gauge field configuration as containing symmetry defects, which correspond to codimesion-1 surfaces in space-time across which the phase associated with the $U(1)$ bundle jumps. This is illustrated in Fig. \ref{fig:symDefects}b. In the present case, we have a phase jump of $\theta_x$ across a plane normal to the $\hat{x}$ direction, and a phase jump of $2\pi y s/L_y$ across a plane normal to the $\hat{t}$ direction. Fig. \ref{fig:symDefects}b illustrates the corresponding diagram for the gauge field configuration $A^{(0)}$.

The gauge field configuration of Eq. \eqref{numeratorGauge} corresponds to an electric field
\begin{align}
E_y = \partial_y A_t - \partial_t A_y = \frac{2 \pi s}{L_y} \delta(t) . 
\end{align}
To evaluate the CS term, we observe that when $A_y = 0$ we can write the CS term as
\begin{align}
  \label{CSeval}
S_{CS} = \frac{1}{2\pi}\int_{M} A_x E_y . 
\end{align}
Thus we get
\begin{align}
  \label{CSfinal}
S_{CS}[A^{(1)}] = s \theta_x . 
\end{align}
Note that if we directly substitute (\ref{numeratorGauge}) into Eq. \eqref{CSresponseAction}, we would naively get $S_{CS} = \frac{1}{4\pi} \int_M A_x \partial_y A_t$, which differs from Eq. \eqref{CSeval} by a factor of $2$. The correct result is Eq. \eqref{CSeval}, as can be verified by considering the fact that the Hall conductivity gives $j_x = \frac{\delta S}{\delta A_x} = \sigma_H E_y = \frac{1}{2\pi} \frac{p}{q} E_y$. The discrepancy arises as there are subtleties in defining $U(1)$ CS theory for non-trivial $U(1)$ bundles \cite{Dijkgraaf1990}. The gauge field configuration we consider has a non-zero electric flux $\int dt dy F_{ty}$ through the $S^2$ parameterized by the $(y,t)$ coordinates. We have circumvented this in the usual way by using a gauge field configuration which is not single-valued (Eq. \eqref{numeratorGauge} is explicitly not periodic in the $y$ direction), but which must be used carefully.  

We thus conclude that in the TQFT,
\begin{align}
\mathcal{Z}(S^2\times S^1; A^{(1)}) = e^{i\theta_x s p/q} \mathcal{Z}(S^2 \times S^1; 0).
  \end{align}
  
We therefore conclude that 
\begin{align}
  \label{TzCS}
  \mathcal{T}(\theta_x, s) \simeq \frac{ \mathcal{Z}(S^2 \times S^1; A^{(1)})}{\mathcal{Z}(S^2\times S^1; A^{(0)})} = e^{i \frac{p}{q} s \theta_x },
\end{align}
where $A^{(1)}$, $A^{(0)}$ are the gauge field configurations of Eq. \eqref{numeratorGauge},\eqref{denominatorGauge}.

Note that since the electric field is not smooth, one will in general expect contributions to the full many-body path integral beyond simply the CS response term, i.e. beyond the contribution of the TQFT. However these additional contributions arise from local contributions to the action involving the field strength and are thus independent of $\oint A_x= \theta_x$. Therefore the proportionality factor in Eq. \eqref{TzCS} is independent of $\theta_x$. Eq. \eqref{TzCS} thus implies that the phase of $\mathcal{T}(\theta_x;s)$ is linear in $\theta_x$, with a slope given by the many-body Chern number $C = s p /q$.

We also note that while the last equality in Eq. \eqref{TzCS} is rigorous in the mathematical formulation of TQFT, the proportionality between the microscopic calculation of $\mathcal{T}(\theta_x;s)$ and the TQFT path integrals relies on the approximation that the TQFT captures the relevant physics. 

Above we considered the case where $M = S^2 \times S^1$. For the purposes of the discussion in the subsequent sections, we note that one can also consider the path integral on $M = T^3$ by treating $(x,y,t)$ as independent periodic coordinates, and the computation of the CS action associated with the gauge field configurations of Eq. \ref{numeratorGauge}, \ref{denominatorGauge} would proceed identically.

\subsubsection{Gauge field configurations related by GL$(2,\mathbb{Z})$}

It will be useful for our later discussion to also consider other gauge field configurations that are related to the current one by changing the cycles along which the symmetry defects have support. Specifically, if we consider the case where the space-time manifold $M =T^3$, we can consider SL$(3,\mathbb{Z})$ coordinate transformations $\vec{x}' = U \vec{x}$, where $\vec{x}$ is the 3-vector $\vec{x}^T = (x,y,t)$ and $U \in \text{SL}(3,\mathbb{Z})$. Under this transformation, the gauge fields transform as derivatives: $\vec{A} = U^T A'$. The CS action is invariant under such a transformation. In the subsequent discussion, we will focus specifically on the case where $y$ is left invariant, but $(x,t)$ transform under an element of SL$(2,\mathbb{Z})$.

More generally, we consider a new gauge field configuration $A'$ such that
\begin{align}
\left(\begin{matrix} \oint_x A' \\ \oint_t A' \end{matrix} \right) = U \left(\begin{matrix} \oint_x A \\ \oint_t A \end{matrix} \right),
\end{align}
where $U \in \text{GL}(2,\mathbb{Z})$. Note that any such $U$ can be decomposed as $U = U_1 U_2$, where $U_2 \in \text{SL}(2,\mathbb{Z})$ and $U_1 =  \left(\begin{matrix} \det U & 0 \\ 0 & 1 \end{matrix}\right)$. We consider the $U_2$ transformation to be obtained by an SL$(2,\mathbb{Z})$ coordinate transformation in the $(x,t)$ space as mentioned above, which keeps the CS action unchanged. We consider the $U_1$ transformation to be obtained by rescaling $\theta_x$ by $\det U$, which has the effect of changing the CS action by a factor of $\det U$. 

\subsection{Space-time surgery and virtual torus through SWAP}
\label{surgerySec}

\begin{figure*}
    \includegraphics[width=.8\textwidth]{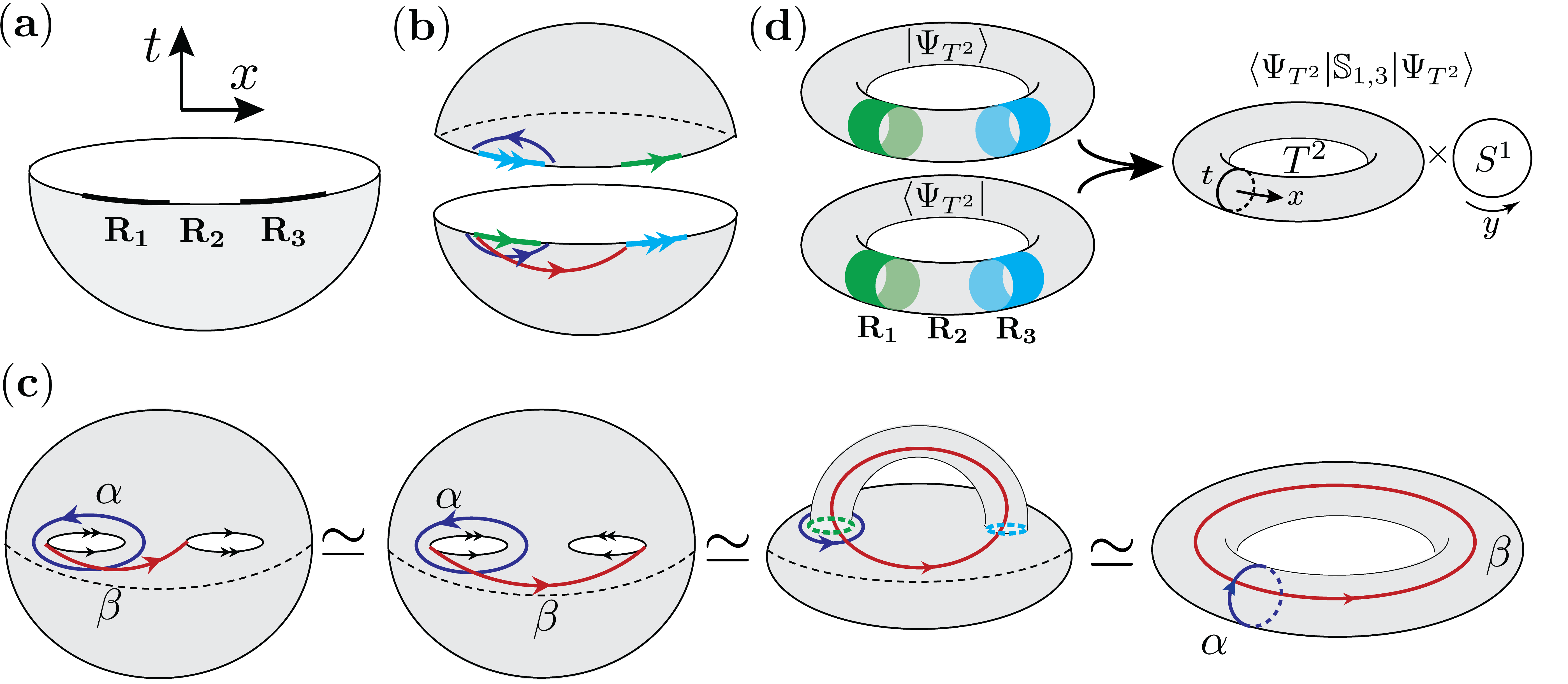}
    \caption{(a) Disk, $D^2$. The state $|\Psi_{\partial D^2}\rangle$ defined on $S^1 = \partial D^2$ in the TQFT can be obtained by evaluating the path integral on a disk. Regions $R_1$, $R_2$, $R_3$ along with the time direction are depicted. (b) Gluing together two disks according to the map $f$ in Eq.~(\ref{fmap}). The gluing scheme is depicted both with the arrow and color scheme. (c) Result of gluing. In the first step one applies a $\pi$-rotation to the right empty region (white) on the sphere. In the second step, identification of the arrows creates a handle. The final object is topologically equivalent to a torus. $\simeq$ are homeomorphisms that continuously deform the manifold. Two non-contractible cycles of the torus ($\alpha, \beta$) are depicted in blue and red. In Sec.~\ref{secSingleCopy}, the symmetry defect operators $V$ and $W$ will have support on the two non-contractible cycles.}
    \label{fig:T2surgery}
\end{figure*}

Here we use the surgery method in TQFT to construct the TQFT path integral on topologically non-trivial space-time manifolds by starting with the state on simple space manifolds. 

\subsubsection{Warmup: (1+1)D TQFT}

For the purpose of our subsequent discussion, we first review how to obtain the path integral of a (1+1)D TQFT on a torus, $T^2$, in terms of the wave function on a circle $S^1 = \partial D^2$, as illustrated in Fig.~\ref{fig:T2surgery}. Let us parametrize space by $x$, from $0$ to $L_x$. We pick three regions, $R_1 = (x_1, x_2]$, $R_2 = (x_2,x_3]$, $R_3 = (x_3,x_4]$, such that $\ell_1 = (x_2-x_1) = (x_4 - x_3) = \ell_3$ and $(x_3 - x_2) = \ell_2$. We then consider the map
\begin{align}
  \label{fmap}
f(x) = \begin{cases}
   x + \ell_1 + \ell_2 &x \in R_1 \\
  x - |\ell_1 + \ell_2| & x \in R_3 \\
  x & x \notin R_1 \cup R_3 \\   
    \end{cases}
\end{align}
The resulting manifold $(\partial D^2) \cup_f (\partial D^2)$ is topologically equivalent to a torus, as illustrated in Fig. \ref{fig:T2surgery}. Note that $f$ in this case is piece-wise continuous, so to obtain the torus we consider smoothing out the singularities at the boundaries of $R_1$ and $R_3$. We therefore conclude that one can obtain the TQFT path integral on a torus through the inner product
\begin{align}
\mathcal{Z}(T^2) = \langle \Psi_{S^1} | \mathbb{S}_{1,3} | \Psi_{S^1}\label{T2fromS1} \rangle,
\end{align}
where $\mathbb{S}_{1,3}$ is the SWAP operation that implements the map $f$ above.

\begin{table*}
\begin{tabular}{ c|c|c|c } 

 \text{\scriptsize{Number of copies}} & \text{\scriptsize{Spatial manifold of}} $|\Psi\rangle$ & \text{\scriptsize{Space-time manifold }$M$ \scriptsize{for}} $\mathcal{Z}(M) = \langle\Psi|\Psi\rangle$  & \text{\scriptsize{Space-time manifold }$M$ \scriptsize{for}} $\mathcal{Z}(M) = \langle\Psi|\mathbb{S}|\Psi\rangle$  \\ 
 \hline\hline 
 \multicolumn{4}{c}{}\\
 \multicolumn{4}{c}{(1+1)D} \\ 
 \hline
 1 & $S^1$ & $S^2$ & $T^2$ (Fig.~\ref{fig:T2surgery}(c), Eq.~\ref{T2fromS1})  \\ 
  \hline
   2 & $S^1$ & $S^2$ & $S^2$ \text{\scriptsize(1-SWAP)} \\ 
    \hline
   2 & $S^1$ & $S^2$ & $T^2$ \text{\scriptsize(2-SWAP)} (Fig.~\ref{fig:bilayerSurgery})\\ 
  \hline
  \multicolumn{4}{c}{}\\
   \multicolumn{4}{c}{(2+1)D, Cylindrical regions $R_i$} \\
   \hline
 1 & $T^2\!=\!S^1\!\times\! S^1$  & $S^2\times S^1$ (Fig.~\ref{fig:symDefects}, Eq.~\ref{eq:Z(S2,S1)}) & $T^3=S^1\times T^2$ (Fig.~\ref{fig:T2surgery}(d), Eq.~\ref{eq:T3fromT2}) \\ \hline
 2  & $T^2$ & $S^2\times S^1$ & $S^1\!\times\! S^2$ \text{\scriptsize(1-SWAP)} (Eq.~\ref{eq:T3fromT2bilayerSingleSWAP})\\ 
 \hline
2  & $T^2$ & $S^2\times S^1$ & $T^3$ \text{\scriptsize(2-SWAP)} (Eq.~\ref{eq:T3fromT2bilayer}) \\
\hline 
\end{tabular}
\caption{\label{tab:surgery} Space-time surgery with SWAP.}
\end{table*}

\subsubsection{Cylindrical regions $R_i$ in (2+1)D}
\label{sec:cylindricalSWAP}

Extending the above arguments to $(2+1)D$, we see that the path integral on the 3-torus, $\mathcal{Z}(T^3)$, can be similarly obtained in terms of the state $|\Psi_{T^2}\rangle$:
\begin{align}
\mathcal{Z}(T^3) = \langle\Psi_{T^2}|\mathbb{S}_{1,3}|\Psi_{T^2}\rangle ,
\label{eq:T3fromT2}\end{align}
where now $\mathbb{S}_{1,3}$ corresponds to the map $(x,y) \rightarrow (f(x), y)$. Note that since $T^2=S^1\times S^1$, the surgery creates a torus from one of the $S^1$ circles, according to Eq.~\ref{T2fromS1} and the other circle simply factors through independently. Therefore, the path integral will be on $T^3=T^2\times S^1$.

\begin{figure*}
    \includegraphics[width=.85\textwidth]{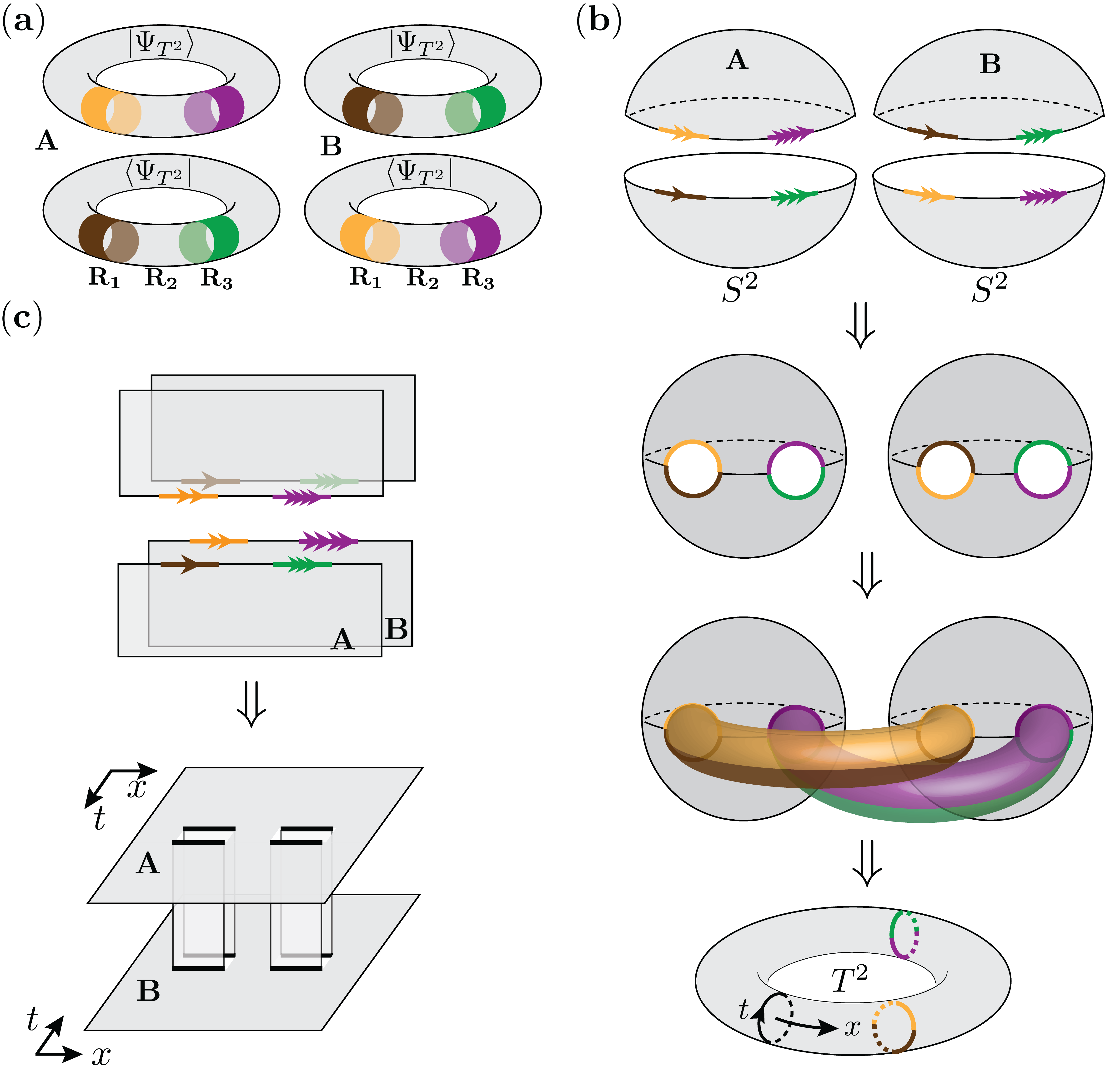}
    \caption{ \textbf{Applying two SWAPs using two copies of the system}
(a) Schematics of of the SWAP operation between two copies, when the wave functions are given on a torus. Gluing scheme is illustrated by gluing like colors together. (b) Since the $y$ direction is an $S^1$, we take a cross-section at a fixed $S^1$. The cross-section consists of two disks $A$ and $B$, which under the SWAPs and inner product get glued to another pair of disks using the gluing map. The result is that the cross-section becomes a torus $T^2$, corresponding to: $\mathcal{Z}(T^2) = \langle \Psi_{S^1}^B | \langle \Psi_{S^1}^A| \mathbb{S}_{R_1^A R_1^B} \mathbb{S}_{R_3^A R_3^B}  |\Psi_{S^1}^A \rangle |\Psi_{S^1}^B\rangle$. (c) Another equivalent illustration  of the gluing process by focusing on a patch of the cross-section corresponding to a fixed choice of $y$.}
    \label{fig:bilayerSurgery} 
\end{figure*}

We note that an alternate way of creating the space-time manifold $T^3$ is to use two copies of the state on $T^2$, as shown in Fig. \ref{fig:bilayerSurgery}, which corresponds to the following inner product in the TQFT:
\begin{align}
  \mathcal{Z}(T^3) = \langle \Psi_{T^2}^B | \langle \Psi_{T^2}^A| \mathbb{S}_{1_A, 1_B} \mathbb{S}_{3_A, 3_B}  |\Psi_{T^2}^A \rangle |\Psi_{T^2}^B\rangle
\label{eq:T3fromT2bilayer}\end{align}
Here we take the regions $R_1^A$, $R_3^A$ (and $R_1^B$, $R_3^B$)  to be defined as in the single layer case. 

We will also note the fact that a single SWAP between two copies of the state on $T^2$ creates a space-time manifold that is topologically equivalent to $S^1 \times S^2$:
\begin{align}
\mathcal{Z}(S^1 \times S^2) = \langle \Psi_{T^2}^B | \langle \Psi_{T^2}^A| \mathbb{S}_{1_A, 1_B} |\Psi_{T^2}^A \rangle |\Psi_{T^2}^B\rangle \label{eq:T3fromT2bilayerSingleSWAP}
\end{align}

The above results are summarized in Table \ref{tab:surgery}.

\subsection{Symmetry defect operators and non-trivial background gauge field configurations}\label{secSymmetryDefect}

At this point, we are ready to combine the results of the previous sections \ref{secCSreponse} and \ref{surgerySec} to obtain our formulas Eqs.~\eqref{T1},\eqref{T2},\eqref{T1U},\eqref{T2UR3},\eqref{T2U}.

So far we have seen how to construct the path integral on various non-trivial space-time manifolds using non-trivial gluing operations corresponding to SWAP between various cylindrical subregions of space. Now we consider obtaining the path integral on these manifolds in the presence of non-trivial gauge field configurations, $\mathcal{Z}(M;A)$ by inserting symmetry defect operators into the expectation values.

\begin{figure*}
    \includegraphics[width=.75\textwidth]{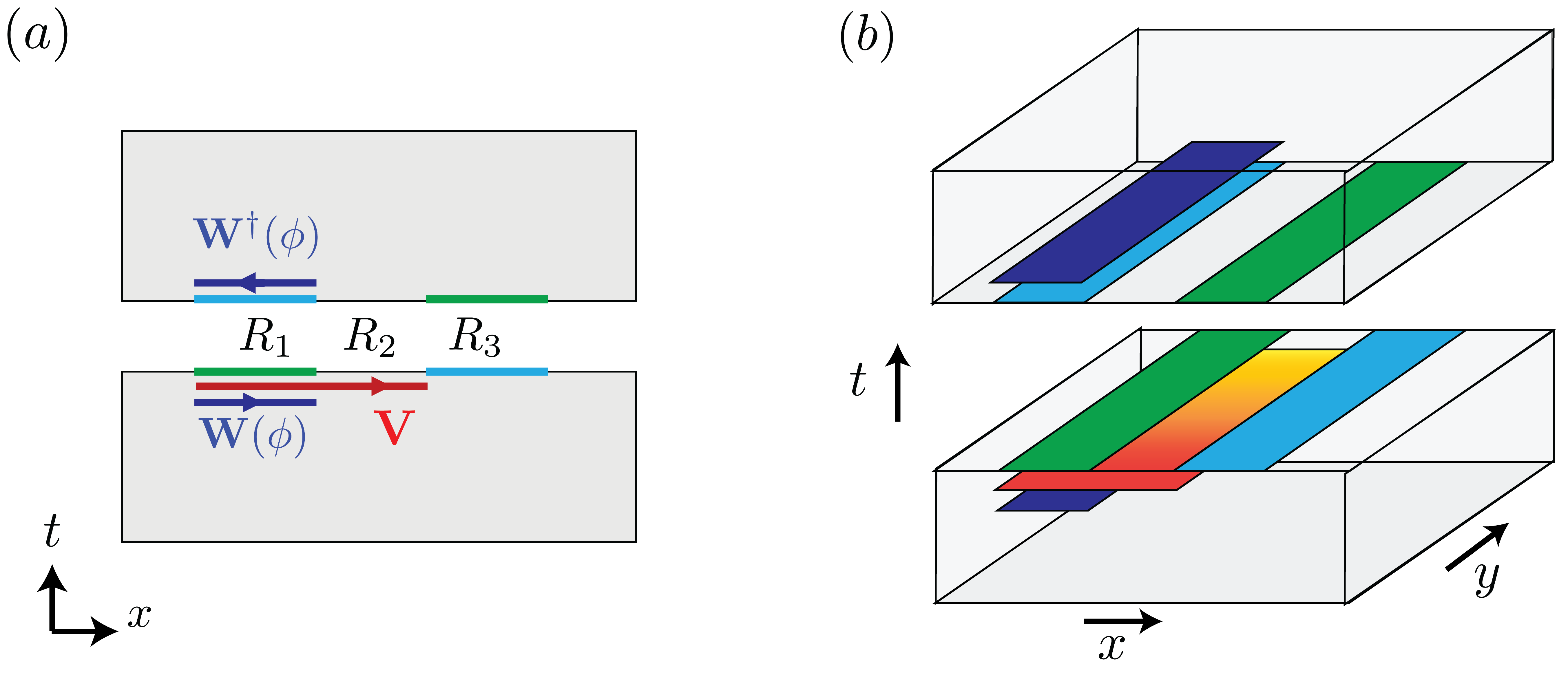}
    \caption{Illustration of symmetry defects for the single copy case. (a) Cross-section of the gluing and symmetry defect configurations, applicable to both the case of cylindrical and rectangular $R_i$. The gluing scheme is depicted with the color scheme. The support of the $W(\phi)$ and $V$ operators becomes topologically equivalent, after gluing, to the $\alpha$ and $\beta$ loops, respectively, shown in Fig. \ref{fig:T2surgery}. (b) Illustration of symmetry defect sheets for the case of cylindrical $R_i$ regions, assuming that the $y$ direction is compactified.}
    \label{fig:oneLayer_symDefect} 
\end{figure*}

\subsubsection{Single layer formula for cylindrical regions $R_i$} 
\label{secSingleCopy}

We start with the case where we obtain the path integral $\mathcal{Z}(T^3)$ from the wave function on a torus $T^2$ by applying a SWAP between two cylindrical subregions $R_1$ and $R_3$, as described in Sec. \ref{sec:cylindricalSWAP}.

We wish to consider a configuration of symmetry defects in the virtual space-time torus which corresponds  to the symmetry defects in Fig. \ref{fig:symDefects}b,c. To obtain this configuration, from Fig. \ref{fig:T2surgery} we see that we should insert symmetry defect operators with support along the $\alpha$ and $\beta$ loops in the $(x,t)$ space and for all $y$. Here the symmetry defect with support along $\alpha$ induces the gauge field configuration of $A_x$ in the notation of Sec. \ref{secCSreponse}. Similarly the symmetry defect with support along $\beta$ induces the gauge field configuration of $A_t$ in the notation of Sec. \ref{secCSreponse}. This configuration of symmetry defects is explicitly illustrated in Fig. \ref{fig:oneLayer_symDefect}.

We are therefore led to the expectation value:

\begin{equation}
  \mathcal{Z}(T^3;A)= 
   \langle \Psi_{T^2} | W_{ R_1}^\dagger(\phi) \mathbb{S}_{1,3} W_{R_1}(\phi) V_{ R_1\cup R_2}^s |\Psi_{T^2} \rangle. 
\end{equation}

Here, $W_{R_1}(\phi)$ creates a phase jump of $\phi$ in the $\hat{t}$ direction (which is normal to $R_1$), while $V_{ R_1 \cup R_2}^s$ creates a phase jump of $2\pi y s/L_y$ in the $\hat{t}$ direction (which is also normal to $R_1\cup R_2$), where $L_y$ is the length of the compactified $y$ direction.

From this, we conclude that we can obtain the many-body Chern number $C$ from a single state in the TQFT 

  \begin{align}
    \label{T2tqft}
e^{i C \phi} &= \frac{\mathcal{Z}(T^3;A^{(1)})}{\mathcal{Z}(T^3;A^{(0)})}  \\
&= \frac{\langle \Psi_{T^2} | W_{ R_1}^\dagger(\phi) \mathbb{S}_{1,3} W_{R_1}(\phi) V_{ R_1\cup R_2}^s |\Psi_{T^2} \rangle}
    { \langle \Psi_{T^2} |W_{ R_1}^\dagger(\phi) \mathbb{S}_{1,3} W_{R_1}(\phi) |\Psi_{T^2} \rangle}.\nonumber
\end{align}

Note that the denominator is real and thus can be ignored when extracting $C$ in terms of the phase of the RHS. 

Furthermore, note that to extract the Chern number, it is sufficient to replace the state on a torus $|\Psi_{T^2}\rangle$ in the above formula with the state (or reduced density matrix) on any region that contains the regions $R_i$. Since all of the operators in the matrix element above have support in region $R_1$, $R_2$, and $R_3$, the properties of the wave function away from $R_1$,$R_2$, $R_3$ are unimportant for extracting the Chern number. It is thus unnecessary to require the $x$ direction to be compactified, so it is sufficient to use the state (or reduced density matrix) on a cylinder, $|\Psi_{S^1 \times I}\rangle$. 

In order to obtain the explicit form of the symmetry defect operators, we must make contact with the microscopic theory from which the TQFT emerges as a long wavelength description. (Strictly speaking, the TQFT does not possess any local operators out of which the symmetry defect operators can be constructed). Since the support of the symmetry defects is orthogonal to the real time direction, these operators can simply be written in terms of the density operator $\hat{n}(x,y)$ in the microscopic theory. The main observation is that inserting $e^{i \int \lambda(x,y) \hat{n}(x,y)}$ into a correlation function at time $t = 0$ has the effect of changing $A \rightarrow A + \lambda(x,y) \delta(t) \hat{t}$. We thus conclude that we can write
\begin{align}
  W_{R}(\phi) &= \prod_{(x,y) \in R} e^{i \phi \hat{n}(x,y) }
  \nonumber \\
  V_{R} &= \prod_{(x,y) \in R} e^{i 2\pi y / \ell_y \hat{n}(x,y) },
\end{align}
where $\ell_y$ is the length of $R$ along the $y$ direction. In the case considered in Eq. \eqref{T2tqft}, $\ell_y = L_y$. We find, therefore, that the Chern number can be obtained from Eq. \eqref{Ceqn}-\eqref{T1}, for the case where the wave function is defined on a torus or cylinder.

In the above discussion, we chose two specific non-contractible cycles, $\alpha$ and $\beta$ which intersect once to arrive at the formula of Eq. \eqref{T2tqft} and \eqref{T1}. In general we could consider any two non-contractible cycles on the torus, related to $(\alpha,\beta)$ by a GL$(2,\mathbb{Z})$ transformation. Specifically, let us consider
\begin{align}
\left( \begin{matrix} \alpha' \\ \beta' \end{matrix} \right) = U \left( \begin{matrix} \alpha \\ \beta \end{matrix} \right) ,
\end{align}
with $U = \left(\begin{matrix} a & b \\ c & d \end{matrix} \right)$, $a,b,c,d \in \mathbb{Z}$. This then leads to the formula
\begin{widetext}
  \begin{align}
    \label{T1UTQFT}
    e^{i (\det U) C \phi}= \left\{ \frac{\langle \Psi_{T^2} | \left( W_{R_1}^{a}(\phi) V_{R_1}^{s c} \right) ^\dagger \mathbb{S}_{1,3} \left( W_{R_1}^{a}(\phi) V_{R_1}^{s c}\right) W_{R_1\cup R_2}^{b}(\phi) V_{R_1\cup R_2}^{s d}| \Psi_{T^2}\rangle}
  {\langle  \Psi_{T^2}| \left( W_{R_1}^{a}(\phi)\right)^\dagger \mathbb{S}_{1,3} W_{R_1}^{a}(\phi) W_{R_1\cup R_2}^{b}(\phi)| \Psi_{T^2}\rangle} \right\},
  \end{align}
  \end{widetext}
  which thus yields Eq. \eqref{T1U}. The appearance of $\det U$ is due to the fact that the CS action changes by a factor of $\det U$ under such a transformation.

  We note that while the denominator $\mathcal{Z}(T^3; A^{(0)}) = 1$ in the TQFT, we find that when evaluating the corresponding expectation values in the full microscopic theory, the denominator is indeed required to obtain correct results. In other words, the correspondence between the expectation values in the microscopic theory and the expectation values in the TQFT holds only for the ratios in the above equations, which is also natural from the form of Eq. \eqref{TaPol}.

\subsubsection{Bilayer formulas for cylindrical regions $R_i$} \label{secTwoCopy}

Following the discussion in Section \ref{surgerySec}, we see that we can also obtain the appropriate gauge field configurations on the space-time manifold $T^3$ by considering two copies of the state. We can read off the support of the symmetry defect operators from Fig. \ref{fig:bilayerSurgery}. The two symmetry defects, $W(\phi)$ and $V$, are inserted in the space-time manifold such that after applying the SWAP operator and forming the virtual torus in the $x-t$ plane, the two symmetry defects will wrap around the non-contractible cycles of the torus. This is explicitly shown in Fig.~\ref{fig:bilayer_defects} which leads to the following formula:
\begin{widetext}
  \begin{align}
    \label{TbilayerTQFT1}
    e^{i C \phi} = \frac{\langle \Psi_{T^2}^B | \langle \Psi_{T^2}^A| W_{ R^A_1}^\dagger(\phi) V_{ R^B_2}^{s\dagger} \mathbb{S}_{1_A,1_B} \mathbb{S}_{3_A,3_B} W_{R^A_1}(\phi) V_{ R^A_2}^s |\Psi_{T^2}^A \rangle |\Psi_{T^2}^B\rangle}
    { \langle \Psi_{T^2}^B | \langle \Psi_{T^2}^A| W_{ R^A_1}^\dagger(\phi) \mathbb{S}_{1_A,1_B} \mathbb{S}_{3_A,3_B} W_{R^A_1}(\phi) |\Psi_{T^2}^A \rangle |\Psi_{T^2}^B\rangle }
\end{align}

\begin{figure*}[t]
    \includegraphics[width=.8\textwidth]{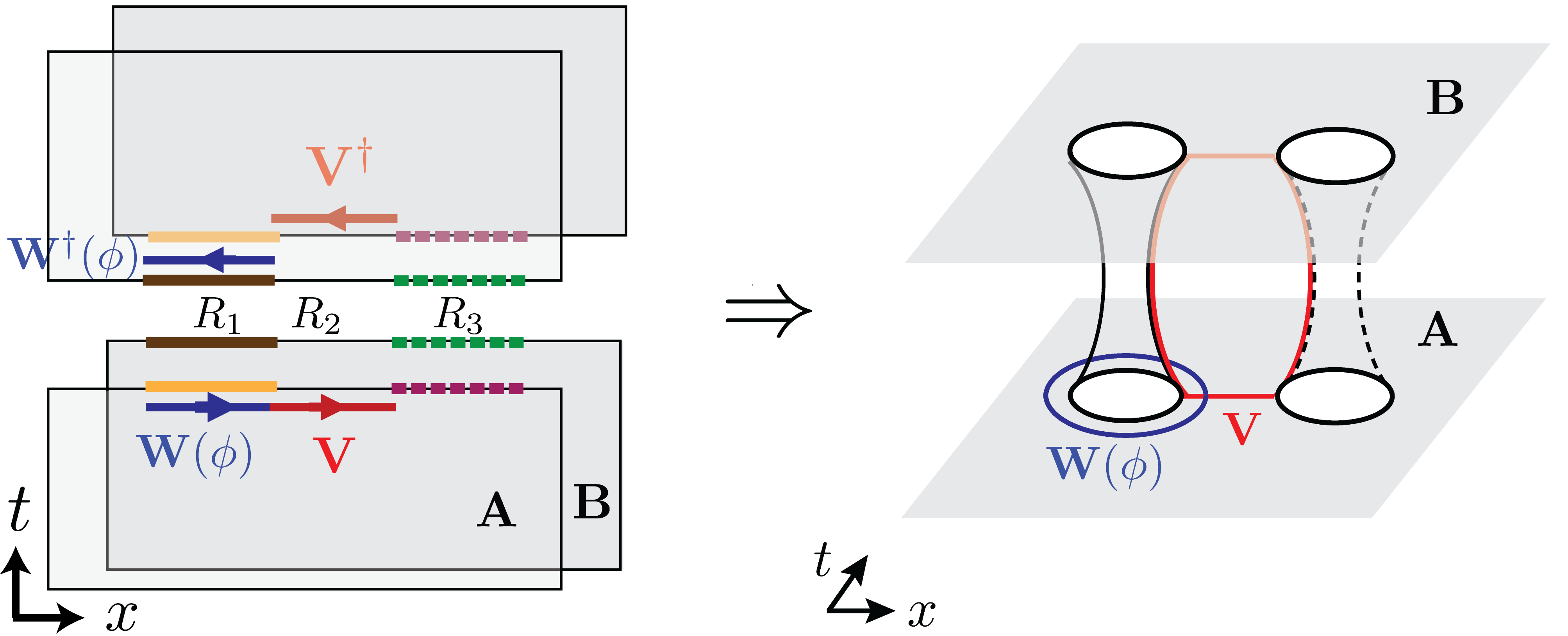}
    \caption{Illustration of symmetry defects for the bilayer case. The SWAP operator and identification is applied to the bra and ket wave functions in the A and B copies of the system. To create a virtual genus in the $x-t$ plane, two SWAP operations should be applied locally between the A and B copies of the wave function in regions $R_1$ and $R_3$. After applying the SWAP operation, the symmetry defects $W(\phi)$ and $V$, will only cross each other in the intersection of region $R_1$ and $R_2$. Consequently, for large system sizes, the SWAP operator in region $R_3$, drawn with dashed lines, is optional and can be removed.}
    \label{fig:bilayer_defects} 
\end{figure*}

Again considering a different set of non-contractible cycles related by $U \in \text{GL}(2,\mathbb{Z})$, we arrive at Eq. \eqref{T2UR3}.

As discussed in Sec. \ref{secCSreponse}, the non-zero contribution to the CS action arises because of the crossing between the two distinct symmetry defects, which here are associated with the $W$ and $V$ operators. Since this crossing occurs away from regions $R_3^A$, $R_3^B$, we can remove $\mathbb{S}_{R_3^A R_3^B}$ from the above formula, to obtain:
  \begin{align}
    \label{TbilayerTQFT2}
  e^{i C \phi} = \frac{\langle \Psi_{T^2}^B | \langle \Psi_{T^2}^A|
  W_{ R^A_1}^\dagger(\phi) V_{ R^B_2}^{s\dagger} \mathbb{S}_{1_A,1_B} W_{R^A_1}(\phi) V_{ R^A_2}^s |\Psi_{T^2}^A \rangle |\Psi_{T^2}^B\rangle }
    { \langle \Psi_{T^2}^B | \langle \Psi_{T^2}^A| W_{ R^A_1}^\dagger(\phi) \mathbb{S}_{1_A,1_B} W_{R^A_1}(\phi) |\Psi_{T^2}^A \rangle |\Psi_{T^2}^B\rangle  }
\end{align}
\end{widetext}

\section{Numerical Results}\label{sec:numerical}

In this section we present extensive numerical simulations of the formulas of Eqs.~\eqref{T1},\eqref{T2},\eqref{T1U},\eqref{T2UR3},\eqref{T2U} for IQH and FQH states. In Section \ref{numericalChernSec} we present our results demonstrating how the Chern number can be extracted using Eqs.~\eqref{T1},\eqref{T2},\eqref{T1U},\eqref{T2UR3},\eqref{T2U} for both cylindrical and rectangular geometries. In Section \ref{sec:stability} we study in detail how the results change when the supports of $V$ and $W$ change in order to better understand the formula and the relation to the TQFT derivation of Section \ref{secTQFT} in terms of crossing of symmetry defects. Finally in Section \ref{sec:scaling} we study the dependence of the magnitude of the expressions in Eqs.~\eqref{T1},\eqref{T2},\eqref{T1U},\eqref{T2UR3},\eqref{T2U} with size of the regions. 

\subsection{Chern number}
\label{numericalChernSec}

Here we provide numerical evidence for the formulas obtained in the previous sections. We consider several integer and fractional quantized Hall bosonic and fermionic states and also a free-fermion Chern insulator state. Using both the single and bilayer formulas for both rectangular and cylindrical geometries, 
we show that our numerical results are consistent with the analytical expectations.

For quantum Hall states, we use matrix product state (MPS) simulations on both cylindrical and rectangular geometries. For bosonic states, we consider the Laughlin state with filling fraction $\nu = 1/2$ \cite{hafezi2007fractional}, the bosonic Jain state at $\nu = 2/3$, the bosonic integer quantum Hall state at $\nu = 2$ \cite{he2017realizing} and the Moore-Read state with $\nu = 1$. For fermions, we consider the integer quantum Hall state with $\nu = 1$, and a free-fermion Chern insulator. For these non-interacting fermionic models, we use the Slater determinant representation of the many-body wave function of the system. This allows us to access larger system sizes compared to the interacting bosonic/fermionic cases. In particular, we use the free-fermion model to investigate the effect of changing the support of the symmetry defect operators on the Chern number in Sec.~\ref{sec:stability}, and to study the scaling of various expectation values with respect to the subregion size in Sec.~\ref{sec:scaling}.

We start with the interacting Hofstadter model on a square lattice, which can realize a variety of fractional quantum Hall states at different values of the filling and flux per plaquette \citep{PhysRevB.14.2239, he2017realizing}. The Hamiltonian is of the form
\begin{equation}
H = -J\sum_{x,y} e^{i2\pi \alpha x}a^{\dagger}_{x,y+1} a_{x,y} + a^{\dagger}_{x+1,y} a_{x,y} + \rm{h.c.},
\label{Hofsdater}
\end{equation}
where $\alpha$ is magnetic flux per plaquette, the sum is taken over $x = 1,\cdots, L_x$ and $y=1,\cdots,L_y$, where $L_x$ and $L_y$ are the number of lattice sites along the $x$ and $y$ directions, respectively. For the bosonic quantum Hall system, with $\nu = 1/2, 2/3$ and $2$, the operator $a(x,y)$ obeys the commutation relation $[a(x,y), a^\dagger(x',y')] = \delta_{x,x'}\delta_{y,y'}$ and the hardcore condition $(a^\dagger(x,y))^2 = 0$. For the Pfaffian MR-state with $\nu = 1$, we impose the three body interaction $(a^\dagger(x,y))^3 = 0$. For the fermionic quantum Hall system, with $\nu = 1$, the operators obey the anti-commutation relation $\{a(x,y), a^\dagger(x',y')\} = \delta_{x,x'}\delta_{y,y'}$. For low flux density $\alpha \ll 1$, the lattice model is known to produce  many-body states similar to the continuum model \cite{hafezi2007fractional}. The ground state is obtained by performing density matrix renormalization group (DMRG) simulation \cite{Cincio2013Characterizing}.

\begin{table}
\begin{tabular}{|c|c|c|c|c|c|}
\hline
Model & s & C & Formula & Geometry & Figure\\
\hline \hline 
QH $\nu=1/2$ B & 2 & 1 & \multirow{4}{*}{Eqs.~(\ref{T1U},\ref{T2UR3})} & \multirow{4}{*}{Cylinder} & \multirow{4}{*}{Fig.~\ref{fig:QH_PBC}}\\ \cline{1-3}
QH $\nu=2/3$ B & 3 & 2 & & & \\ \cline{1-3}
QH $\nu=2$ B & 1 & 2 & & & \\ \cline{1-3}
QH $\nu=1$ F & 1 & 2 & & & \\ \cline{1-3}
QH $\nu=1$ B MR & 2 & 2 & & & \\ \cline{1-3}
\hline
QH $\nu=1/2$ B & 2 & 1 & \multirow{2}{*}{Eqs.~(\ref{T1U},\ref{T2UR3})} & \multirow{2}{*}{Rectangle} & \multirow{2}{*}{Fig.~\ref{fig:QH_OBC}}\\ \cline{1-3}
QH $\nu=1$ F & 1 & 1 & & & \\ \cline{1-3}
\hline
QH $\nu=1/2$ B & 2 & 1 & \multirow{6}{*}{Eq.~(\ref{T2U})} & \multirow{4}{*}{Cylinder} & \multirow{6}{*}{Fig.~\ref{fig:two_copy_no_R3}}\\ \cline{1-3}
QH $\nu=2/3$ B & 3 & 2 & & & \\ \cline{1-3}
QH $\nu=2$ B & 1 & 2 & & & \\ \cline{1-3}
QH $\nu=1$ F & 1 & 1 & & & \\ \cline{1-3}
QH $\nu=1$ B MR & 2 & 2 & & & \\ \cline{1-3}
QH $\nu=1/2$ B & 2 & 1 &  & \multirow{2}{*}{Rectangle} & \\ \cline{1-3}
QH $\nu=1$ F & 1 & 1 & & & \\ \cline{1-3}
\hline
\multirow{1}{*}{Chern insulator} & \multirow{1}{*}{1} & \multirow{1}{*}{1} & \multirow{1}{*}{Eq.~\eqref{T1U}} & Cylinder & \multirow{1}{*}{Fig.~\ref{fig:Chern_vs_Vsupport}}
 \\
\hline
\end{tabular}
\caption{\label{tab:numerics} Summary of the numerical simulations. B/F stands for bosonic/fermionic states, respectively. MR stands for Moore-Read state. }
\end{table}

\begin{figure}[h]
    \includegraphics[width=.5\textwidth]{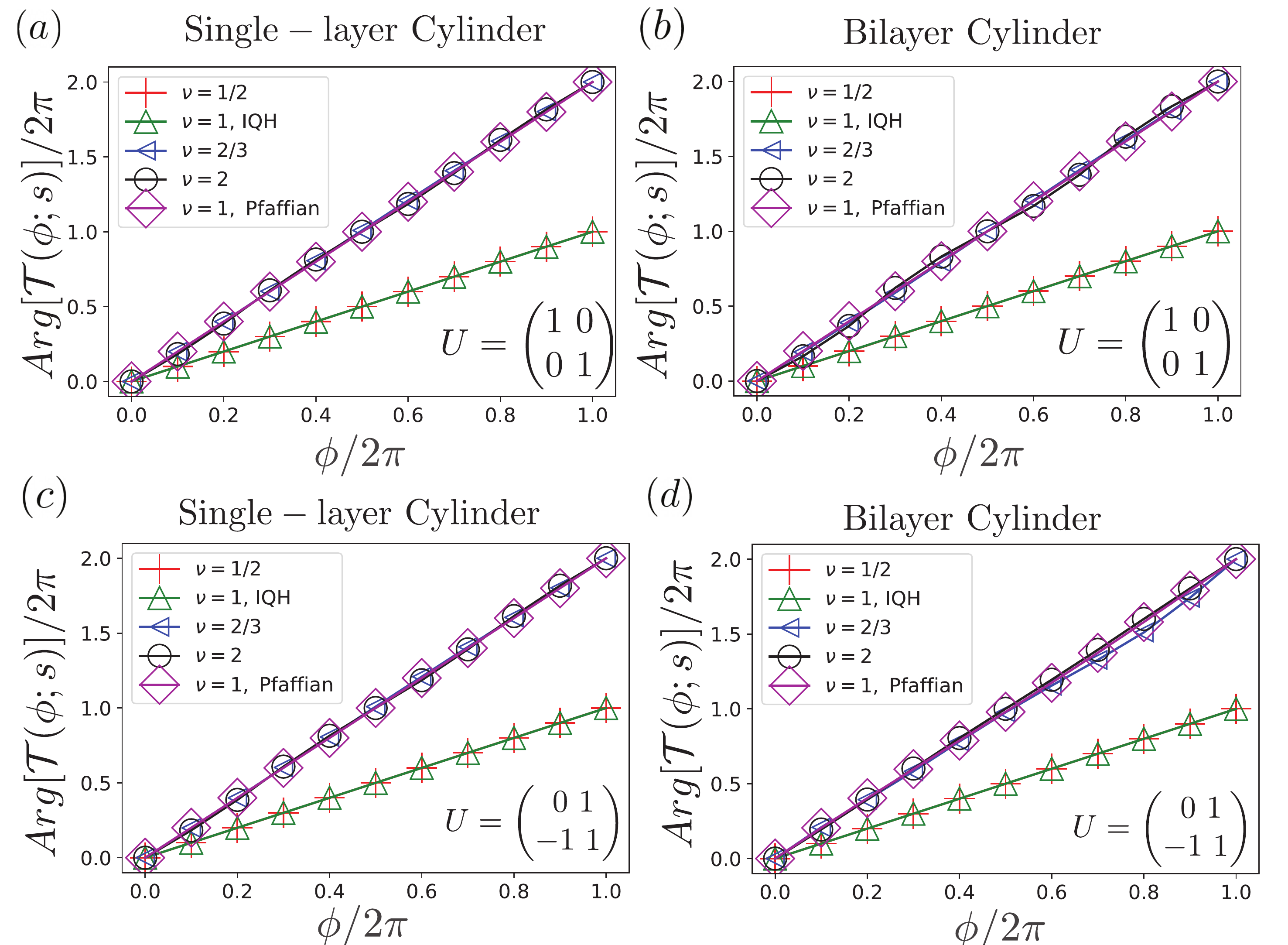}\\
    \caption{ The simulation results for single and bilayer formulas (Eqs.~\eqref{T1U} and \eqref{T2UR3}) with cylindrical geometry for quantum Hall states. For the bosonic Laughlin state with filling fraction $\nu = 1/2$, we choose $\alpha = 1/6$, $L_y = \ell_1 = \ell_2 = \ell_3 = 6$, $L_x = 30$ and $s = 2$. For the Jain sequence state with $\nu = 2/3$, $\alpha = 1/9$, $L_y = \ell_1 = \ell_2 = \ell_3 = 9$, $L_x = 45$ and $s = 3$. For the bosonic integer quantum Hall state with $\nu = 2$, $\alpha = 1/6$, $L_y = \ell_1 = \ell_2 = \ell_3 = 6$, $L_x = 40$ and $s = 1$. For the fermionic integer quantum Hall state with $\nu = 1$, $\alpha = 1/6$, $L_x = 40$, $L_y =  \ell_1 = \ell_2 = \ell_3 = 6$ and $s = 1$. For the bosonic Pfaffian MR-state with $\nu = 1$, $\alpha = 1/6$, $L_x = 60$, $L_y = 6, \ell_1 = \ell_2 = \ell_3 = 12$ and $s = 2$. Corresponding $U$'s are shown in each panel.}
    \label{fig:QH_PBC}
\end{figure}

We first consider a cylindrical geometry, where we impose periodic boundary conditions along the $y$ axis of Fig.~\ref{fig:spatialTopo}(a) and choose the regions $R_i$ to wrap the $y$ direction, such that $\ell_y = L_y$. The numerical results for $\arg[\mathcal{T}(\phi; s)]$ as defined in Eq. \eqref{T1U} and \eqref{T2UR3}, along with the choice of parameters in the simulation, are shown in Fig. \ref{fig:QH_PBC}. As we can see in this figure, if the system size is sufficiently large, the slopes of the curves are nearly constant and thus can be used to obtain the Chern number. 

\begin{figure}[h]
    \includegraphics[width=.5\textwidth]{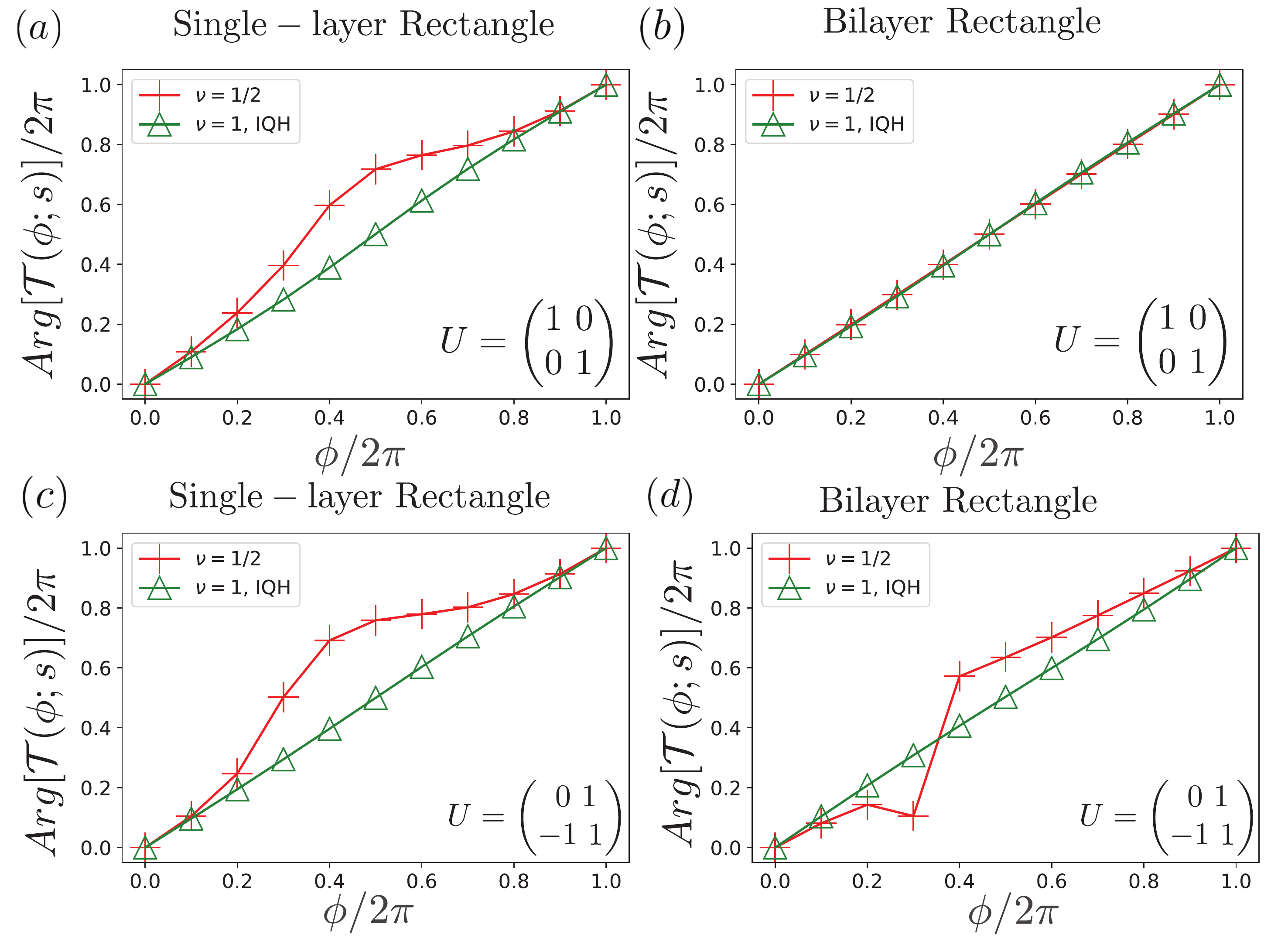}\\
    \caption{The simulation results for single and bilayer formulas (Eqs.~\eqref{T1U} and \eqref{T2UR3}) with rectangular geometry for quantum Hall states. For the bosonic Laughlin state with $\nu = 1/2$, we choose $\alpha = 1/6$, $L_x=40$, $L_y = 9$,  and $\ell_1 = \ell_2 = 9$, $\ell_y = 7$ and $s = 2$. For the fermionic integer quantum Hall state with $\nu = 1$, $\alpha = 1/4$, $L_x = 30$, $L_y = 8$, $ \ell_1 = \ell_2 = \ell_y= 8$ and $s = 1$. Corresponding $U$'s are shown in each panel.}
    \label{fig:QH_OBC}
\end{figure}

\begin{figure}[h]
    \includegraphics[width=.5\textwidth]{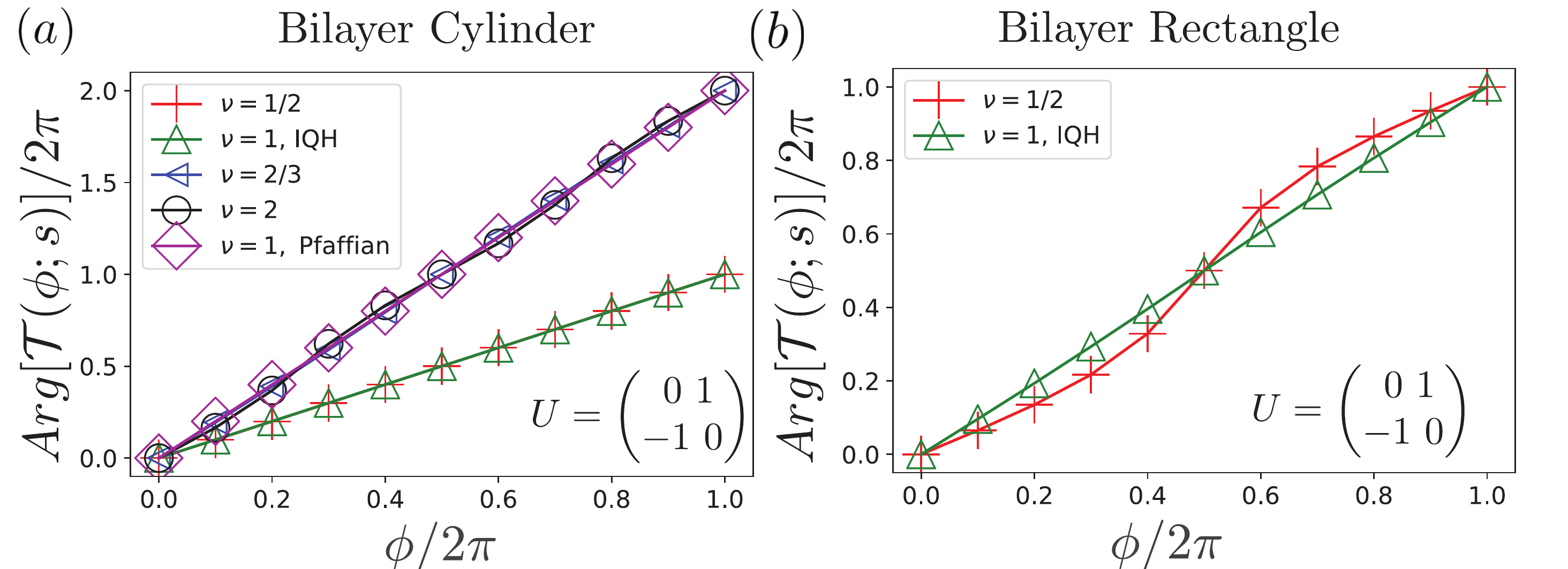}\\
    \caption{The simulation result of single SWAP for bilayer (Eq.~\eqref{T2U}). (a) Cylindrical geometry, with layer parameters identical to Fig.~\ref{fig:QH_PBC}. (b) Rectangular geometry, with  layer parameters identical to Fig.~\ref{fig:QH_OBC}.}
    \label{fig:two_copy_no_R3}
\end{figure}

We also consider systems in the rectangular geometry as shown in Fig. \ref{fig:spatialTopo}(b),(c). The length of the swapped regions $R_i$ along the $y$ direction is denoted by $\ell_y$ while the total length along the $y$ direction is denoted by $L_y$ as before. Since the Landau level of the Hofstadter model is not flat for small system sizes in the rectangular geometry, we do not find robust FQH states at $\nu = 2/3$ and $\nu = 2$ at the system sizes that we access. The winding number of $\arg[\mathcal{T}(\phi; s)]$ is an integer by definition, however it jumps to different values and is not converged with system size for $\nu = 2/3$ and $\nu = 2$. This is also the case even when we flatten the bands using the Kapit-Mueller tunneling terms \cite{kapit2010exact}.  We thus present in Fig.~ \ref{fig:QH_OBC} the result of the Chern number calculation for the $\nu = 1/2$ Laughlin state and the fermionic integer quantum Hall state with $\nu = 1$, which are robust to boundary effects in our simulations.  As expected the winding number of the phase of the twist operator generates the correct Chern number. However, unlike in Fig.~\ref{fig:QH_PBC}, where we observe a linear behavior in $\phi$, the finite size effects are significantly larger; nevertheless the winding number still correctly reproduces the Chern number in the system sizes that we have accessed.

\begin{figure}[h]
    \includegraphics[width=.5\textwidth]{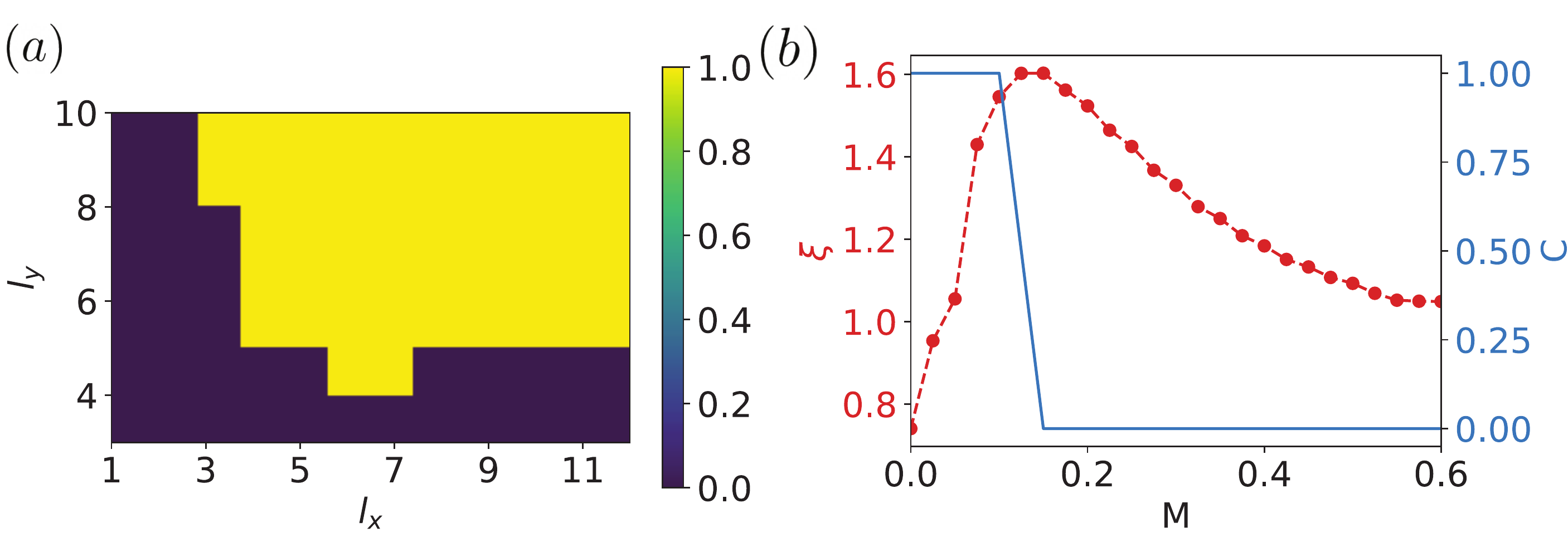}\\
    \caption{(a) The simulation result for bilayer formula Eq. \eqref{T2U} with various region size. We consider the Laughlin state with $\nu = 1/2$, we choose $\alpha = 1/6$, $L_x = 40$, $L_y = 12$ and $\ell_1 = \ell_2 = \ell_x$ and $U = \begin{pmatrix}0~1\\-1~0\end{pmatrix}$. (b) Phase transition of a  Laughlin state to a Mott insulator. The correlation length $\xi$ (red) and the Chern number $C$ (blue) are shown as functions of $M$. $M$ is defined in the main text. We choose $L_x = 25$, $L_y = 12$, $\alpha = 1/6$ and $U = \begin{pmatrix}0~1\\-1~0\end{pmatrix}$. The correlation length remains finite around the critical point due the truncation of the bond dimension.}
    \label{fig:QH_rectangle}
\end{figure}

In Fig.~\ref{fig:two_copy_no_R3}, we present the results for the bilayer formula where the SWAP in region $R_3$ is removed, as in Eq.~\eqref{T2U}. Although both the single and two SWAP bilayer formulas of Eq. \eqref{T2UR3} and \eqref{T2U} can be used to obtain the Chern number, the single SWAP formula Eq. \eqref{T2U} is significantly easier to implement in an experiment, and therefore, this formula is used in Ref. \cite{ourprl}. The results are consistent with the theoretical prediction. We summarize the simulation results for Figs. \ref{fig:QH_PBC}, \ref{fig:QH_OBC} and \ref{fig:QH_rectangle} in Table \ref{tab:numerics}.

In order to provide more evidence for the formulas of Eqs.~\eqref{T1},\eqref{T2},\eqref{T1U},\eqref{T2UR3},\eqref{T2U} in the rectangular geometry, we perform finite size scaling for Eq. \eqref{T2U}, as shown in Fig. \ref{fig:QH_rectangle}(a). We observe that when the areas of the regions $R_1$ and $R_2$ are large enough, the resulting Chern number converges to one. In addition, we also consider a fractional quantum Hall to Mott insulator phase transition by adding an extra potential term $V = M\sum (-1)^{p_x+y}\hat{n}(x,y)$, where $p_x = \lfloor \alpha x \rfloor$ in Eq. \eqref{Hofsdater}. The system undergoes a phase transition from a FQH phase to a Mott insulator when $M$ is increased \cite{hafezi2007fractional}. In Fig. \ref{fig:QH_rectangle}, we plot the correlation length $\xi$ and the Chern number computed by Eq. \eqref{T2U} as functions of $M$. Although the correlation length peaks around the critical point ($M \sim 0.2$), it remains finite due to the finite bond dimension of the MPS which is chosen to be $\chi = 200$. We can see that the value of $M$ at which the phase transition in the Chern number occurs and the the value of $M$ at which the extremum of the correlation length occurs, are approximately coincident. This coincidence, which by increasing the bond dimension becomes more accurate, indicates that the SWAP formula in Eq. \eqref{T2U} can still detect the Chern number correctly, even in  rectangular geometries.

We have also numerically evaluated with our DMRG simulations both the single layer and bilayer formulas with cylindrical and rectangular geometry, Eq. \eqref{T1U} and \eqref{T2UR3}, with all possible choices of $U \in \text{SL}^{\pm}(2,\mathbb{Z})$ that have $|a|, |b|, |c|, |d| \leq 1$. While cases with $|a| > 1$ and $b=0$ trivially correspond to rescaling $\phi$, in general choosing $|a|, |b|, |c|, |d| > 1$ does not typically give stable results at the system sizes that are accessible to us. 

\subsection{Stability under variation of the support of the symmetry defects}
\label{sec:stability}

In this subsection, we study the stability of the SWAP formulas obtained for the Chern number under the variation of the support of the $V$ and $W(\phi)$ symmetry defect operators with respect to each other. 
In order to numerically study this problem, we need to consider larger systems compared to the cases considered in Fig. \ref{fig:QH_PBC}, \ref{fig:QH_OBC}, \ref{fig:two_copy_no_R3}. 

Thus, instead of studying FQH systems, we consider a half-filled spinless free-fermion Chern insulator, which can be studied on larger lattices compared to FQH systems. As before, we can consider both cylindrical and rectangular geometries in the $x-y$ plane. The momentum space Hamiltonian of this system on a square lattice is described by 
\begin{align}
    H = \sin k_x\sigma_x + \sin k_y\sigma_y + (M_c + 2 - \cos k_x-\cos k_y)\sigma_z, \label{chernInsH}
\end{align}
where $\sigma_i$'s denote the Pauli matrices on a Hilbert space with two orbitals per site, and $(k_x, k_y)$ belong to a square lattice Brillouin zone (BZ), $k_i \in (-\pi, \pi)$. For $-2<M_c<0$ and $-4<M_c<-2$, this Hamiltonian has a Chern number $C=-1, +1$, respectively. 

\begin{figure}[h]
\includegraphics[width=.5\textwidth]{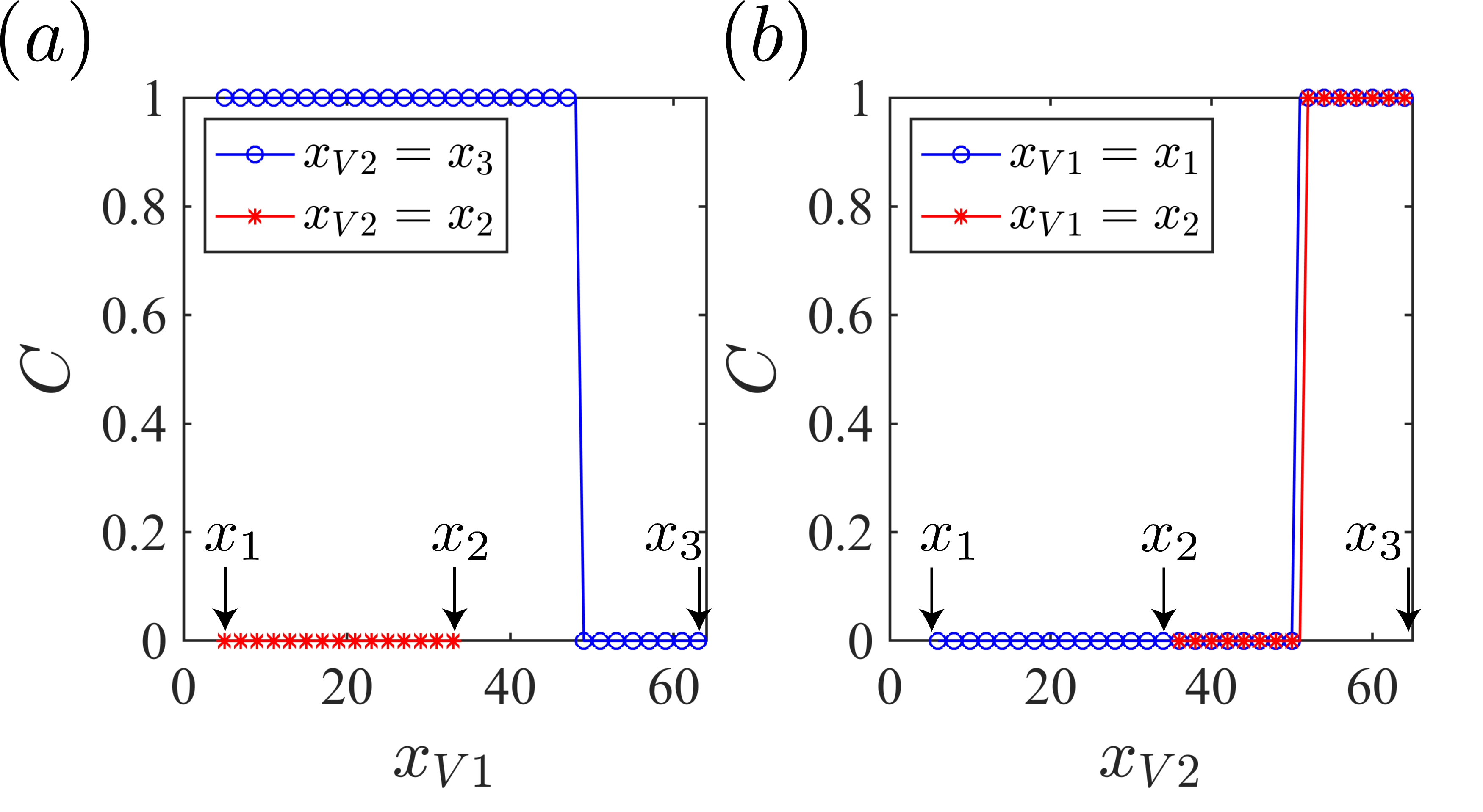}\\
\caption{Chern number extracted from Eq. \eqref{T1VaryVSupp} as a function of the limits of the support of the $V$ operator along the longitude of a cylinder which are denoted by $x_{V1}$ and $x_{V2}$. We consider the Chern insulator model introduced in Eq.\eqref{chernInsH} with $M_c=-3$, on a cylindrical geometry with $L_x =100, L_y=10$, and $\ell_1 = \ell_2 = 30, \ell_y =10$. Here, we choose $x_1=5, x_2=35, x_3=65$. The longitudinal limits of the regions $R_1, R_2$ and  $R_3$, are determined by $5\leq x<35$, $35\leq x < 65$, and $65 \leq  x<95$, respectively. (a) The blue and red curves correspond to $x_{V2}=x_2$ and $x_{V2} =x_3$, respectively. By increasing $x_{V1}$, in the red curve, $R_V$ can cover regions $R_1$ and $R_2$, while in the blue curve, $R_V$ is only allowed to cover the region $R_1$. (b) The blue and red curves correspond to $x_{V1}=x_1$ and $x_{V1} =x_2$, respectively. By increasing $x_{V2}$, in the blue curve, $R_V$ can cover regions $R_1$ and $R_2$, while in the blue curve, $R_V$ is only allowed to cover the region $R_2$.} \label{fig:Chern_vs_Vsupport}
\end{figure}

For concreteness, we use the above model to implement Eq. \eqref{T1} on a cylinder with cylindrical regions $R_i$. We fix the support of the operator $W(\phi)$ to be equal to the region $R_1$, and let the support of the operator $V$ to be varied,
\begin{align}
    \label{T1VaryVSupp}
    \mathcal{T}(\phi;s) = \langle 0| W ^\dagger_{R_1}(\phi) \mathbb{S}_{1,3} W_{R_1}(\phi) V^s_{R_V}|0\rangle ,
\end{align}
where $R_V$ denotes the support of the operator $V$, which is supposed to be cylindrical as well. We introduce $x_{V1}$ and $x_{V2}$ to determine the support of the $R_V$ region along the $x$ direction according to,
\begin{align}
R_V = \{(x,y) | x_{V1} \leq x < x_{V2}, 0 \leq y \leq L_y \},
\end{align}
where now $\ell_y = L_y$, the length of the cylinder along the $y$ direction. To study the stability of the winding number in Eq. \eqref{T1VaryVSupp}, we change $x_{V1}$ and $x_{V2}$ separately. Thus, we consider two different schemes where we first fix $x_{V2}$ and vary $x_{V1}$, and then fix $x_{V1}$ and vary $x_{V2}$. The corresponding numerical results are depicted in Fig. \ref{fig:Chern_vs_Vsupport}(a), and (b), respectively. In Fig. \ref{fig:Chern_vs_Vsupport}(a), the two curves correspond to two different values of $x_{V2}$. In the red curve by keeping $x_{V2}$ constant at $x_2$, we are allowed to only cover the region $R_1$, while in the blue curve by fixing $x_{V2}=x_3$, we can cover both regions of $R_1$ and $R_2$. Here, we notice that when the support of the $V$ operator is only limited to the region $R_1$, we cannot detect the Chern number correctly. This implies that the support of $V$ in region $R_1$ is not as crucial as its support in region $R_2$. This observation is further supported by considering Fig. \ref{fig:Chern_vs_Vsupport}(b) where we vary $x_{V2}$ and hold $x_{V1}$ constant at $x_1$ and $x_2$. In the former case which corresponds to the blue curve, $R_V$ can include both regions of $R_1$ and $R_2$, while in the latter case which corresponds to the red curve, $R_V$ can only cover the region $R_2$. Here, we observe that when $x_{V2}$ is sufficiently increased to allow $R_V$ have a substantial overlap with the region $R_2$, the winding becomes non-zero. Since the two curves almost overlap, we can again infer that to detect the Chern number from Eq. \eqref{T1VaryVSupp}, we only need to cover the region $R_2$ in the support of $V$, and including the region $R_1$ only plays a minor role in stabilizing the results.

\begin{figure}[h]
\includegraphics[width=.5\textwidth]{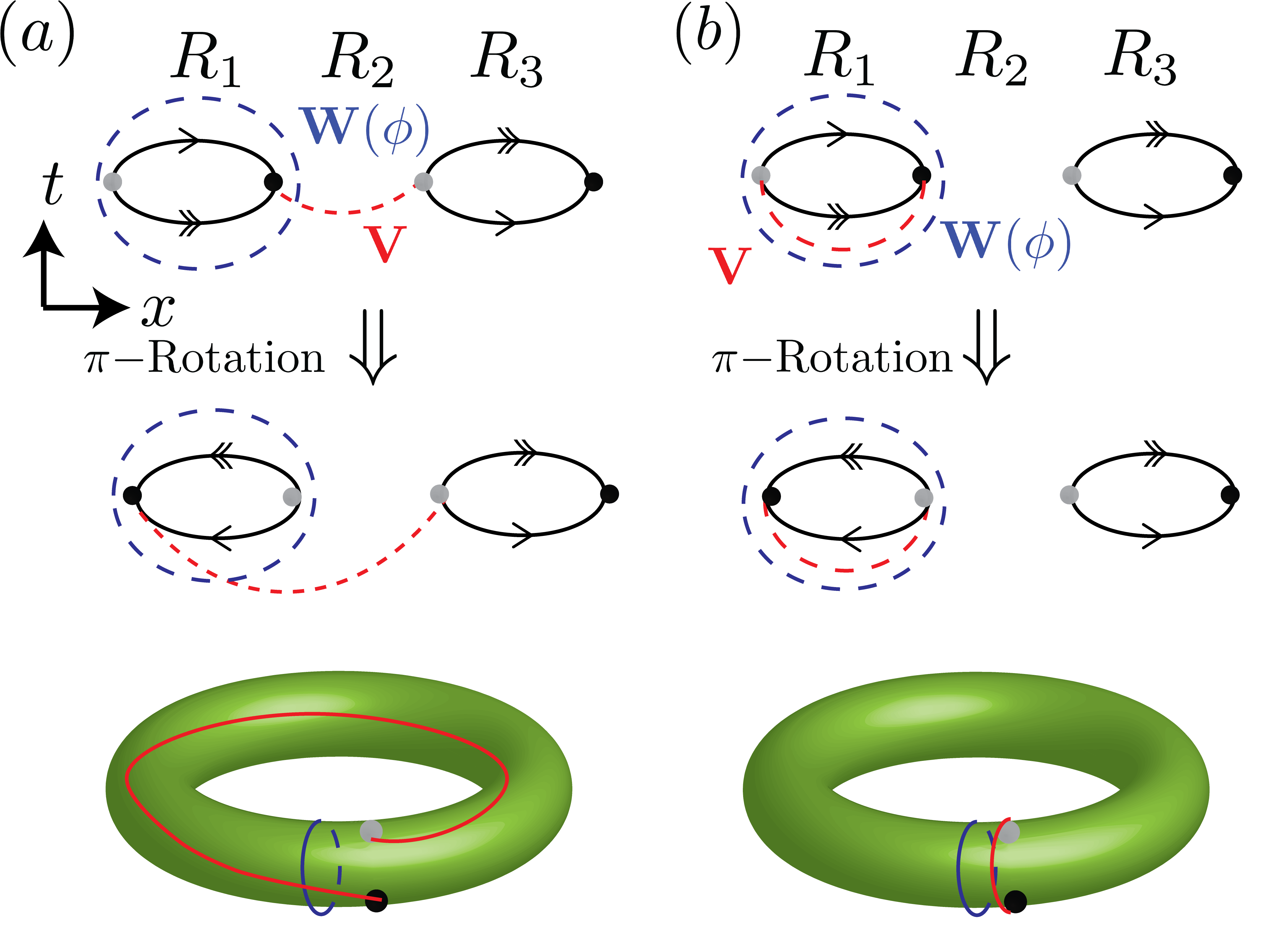}\\
\caption{Configuration of the symmetry defects, $V$ and $W(\phi)$ in the $x-t$ plane, when the support of $V$ is varied. For ease of visualization, we choose the support of $W(\phi)$ to be slightly larger than the region $R_1$. After applying a $\pi$-rotation to the left circle in the second row, the support of $V$ is changed accordingly. In the third row the corresponding torus formed by identification of the arrows is depicted. The symmetry defects associated with $V$ and $W(\phi)$ are shown with a red and blue curve, respectively. (a) The support of $V$ encompasses the region $R_2$. After swapping and identifying the regions with the same number of arrows, the symmetry defects can still cross each other. (b) The support of $V$ encompasses the region $R_1$. After swapping and identifying the regions with the same number of arrows, the symmetry defects no longer cross each other.} \label{fig:varyVsupp_SymDef}
\end{figure}

This observation can be explained from a TQFT perspective. To do this, we investigate the crossing of the symmetry defects when $R_V$ is allowed to change along the longitude of the cylinder. In particular, we study the crossing of the symmetry defects after applying the SWAP operator. This is demonstrated in Fig. {\ref{fig:varyVsupp_SymDef}}, where in the left panel $R_V$ encompasses the region $R_2$, while in the right panel $R_V$ encloses the region $R_1$. Note that in both cases the support of $V$ does not form a closed loop. To aid visualization, in the second row we apply a $\pi$-rotation to the circles on the left-hand side. After identification of the regions $R_1$ and $R_2$ in the forth row, we see that  while in (a) the two symmetry defects can still cross each other, in (b) the two symmetry defects become parallel to each other. Note that this observation is in accordance with our previous Chern-Simons description. This is because from the Chern-Simons theory in order to extract the Chern number we only need to ensure that the two symmetry defects in the $x-t$ plane cross each other. Therefore, enforcing the two symmetry defects to wrap around the two non-contractible cycles of the space-time torus should be only considered as an extra measure to guarantee that the crossing occurs and the results are stable. Before ending, we also point out that due to the finite correlation lengths of the systems in our simulations, in Fig. \ref{fig:Chern_vs_Vsupport} even before encompassing the region $R_2$ fully, we are able to acquire a non-vanishing winding number for $\mathcal{T}(\phi;s)$.

\subsection{Magnitude of the SWAP Expressions}\label{sec:scaling}

An important aspect of the expressions in Eqs.~\eqref{T1},\eqref{T2}, \eqref{T1U},\eqref{T2U} is that while the phase can be used to obtain the Chern number, the magnitude of each individual expectation value in either the numerator or the denominator is exponentially small in the size of the subregions involved in the SWAP operator. For example, in the absence of any symmetry, the reduced density matrices $\rho_{R_1}$ and $\rho_{R_3}$ may be different; thus an expectation value involving the SWAP operator $\mathbb{S}_{1,3}$ will generally be exponentially small in the area of $R_1$ and $R_3$. While the phase is still well-defined and the Chern number can be extracted in principle for arbitrarily large $R_i$, the exponential decay of the amplitude means that in practice for large enough $R_i$ the phase cannot be reliably distinguished from numerical or experimental error. Therefore, in practice for numerical or experimental studies, we need to consider sizes of $R_i$ that are larger than the correlation length but small enough that the phase can be reliably extracted.

We empirically find that the ratio of the numerator and denominator in the expressions of Eq. \eqref{T1U}, \eqref{T2U} has magnitude of order one. However, it is not clear whether there is a useful method to directly calculate the ratio without calculating the expectation values in the numerator and denominators individually. 

If the system has translation symmetry so that the reduced density matrices in the two regions are equal, $\rho_{R_1} = \rho_{R_3}$, then the magnitude of the expectation values will no longer decay exponentially with the area of $R_i$, but rather with the perimeter of $R_i$. This follows from the area law of the entanglement entropy, since the SWAP operator effectively changes the entanglement structure along the perimeter of $R_1$ and $R_3$. 

In this section, we study in more detail the magnitude of the expectation values in the translationally invariant case in order to better understand how it varies with the size of $R_i$. We expect that for the rectangular geometry this exponential suppression is proportional to $\ell_y+\ell_1$ while for cylindrical geometries where the $y$-dimension is compactified, it will be proportional to $L_y = \ell_y$. The cylindrical geometry is the case that we will study for our numerical studies in this subsection. 

We first report based on our numerical studies that the flux angle $\phi$ does not significantly change the system size dependence of the magnitude of the expectation values in $\mathcal{T}(\phi; s)$. Thus, we focus on the case $\phi = 0$.

Next, we note that in our single layer formula, Eq.~\eqref{T1U}, $V$ generally has support both in region $R_1$ and $R_2$, for most choices of the matrix $U$. Therefore, as a generic example of these formulas, we consider the single-copy formula as presented in Eq.~\eqref{T2tqft} where $V$ has support in both regions $R_1$ and $R_2$. Notice that when $\phi=0$, the numerator is $\langle \mathbb{S}_{1,3} V_{R_1\cup R_2}^s \rangle$ and the denominator is $\langle \mathbb{S}_{1,3} \rangle $. Studying the numerator and denominator separately thus also allows us to distinguish the effects of the SWAP operator from $V_{R_1\cup R_2}$ in our formulas. 

In order to study the scaling behavior of the formulas for a wide range of system sizes, instead of using a FQH system, we consider the half-filled free-fermion Chern insulator introduced in Eq.~ \eqref{chernInsH}, which we study in the cylindrical geometry. 

\begin{figure}[h]
\includegraphics[width=.5\textwidth]{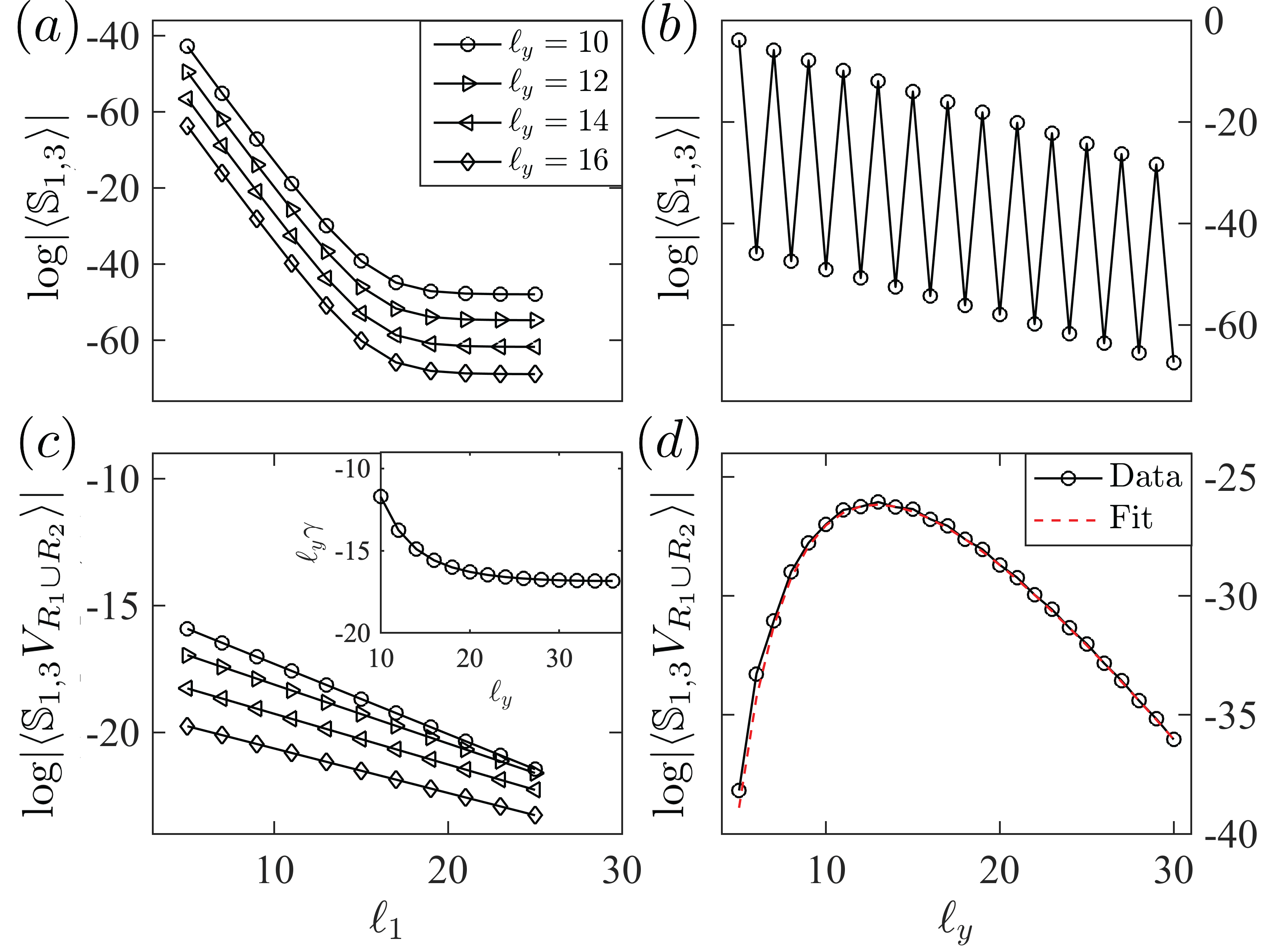}\\
\caption{System size dependence of the magnitude of the numerator and denominator of Eq. \eqref{T2tqft} for a half-filled Chern insulator. In the top panel the logarithm of the SWAP operator's amplitude, $\log |\langle\mathbb{S}_{1,3}\rangle|$, which corresponds to the denominator of \eqref{T2tqft} at $\phi=0$, is plotted as a function of $\ell_1$, and $\ell_2$ in (a) and (b), respectively. (a) Different curves belong to different $\ell_y$'s. (b) The behavior of the SWAP operator is displayed as a function of $\ell_y$. In the bottom panels the amplitude of the product of the SWAP, $\mathbb{S}_{1,3}$, and the exponentiated polarization operators, $V_{R_1\cup R_2}$, is plotted as a function of $\ell_1$, and $\ell_y$ in (c) and (d), respectively. (c) Different curves correspond to different $\ell_y$'s. The same marker symbols as in (a), are used in (c). (d) The behavior of the numerator is displayed as a function of $\ell_y$. The black solid line depicts the results of the simulations, and the red dashed line is a minimal polynomial fit based on Resta's polarization argument using only $\ell_y^{m}$, with $m=\{-1, 0, 1\}$.}
\label{fig:amp_scaling}
\end{figure}

Let us first consider $|\langle \Psi | \mathbb{S}_{1,3} | \Psi\rangle|$. As discussed above, we expect on general grounds that on a cylinder $\log |\langle \Psi | \mathbb{S}_{1,3} | \Psi\rangle| \propto \ell_y$. Our numerical results are depicted in Fig. \ref{fig:amp_scaling}(a) as a function of $\ell_1$ when $\ell_2$ is a constant. A similar behavior is observed in the reverse situation when $\ell_2$ varies and $\ell_1$ is a constant. The asymptotic behavior of $\log |\langle \mathbb{S}_{1,3} \rangle|$ is independent of $\ell_1$. In our simulations the asymptotic regime sets in when $\ell_1$ becomes comparable to $\ell_2$. Furthermore, we verify that this asymptotic value decreases exponentially as $\ell_y$ increases. This change of behavior is more carefully studied in Fig.~\ref{fig:amp_scaling}(b) where in addition to an exponential decay in $\ell_y$ we also observe an oscillating behavior with a periodicity of $2$ lattice constants. While the exponential decay in $\ell_y$ as explained before comes from the disruption of the entanglement at the boundaries of the swapped regions, this extra oscillating behavior is analogous to Friedel oscillations of the Renyi entropy due to the Fermi surface \cite{swingle2013oscillating}.

Next we study the length dependence of the amplitude $|\langle \mathbb{S}_{1,3} V_{R_1 \cup R_2}\rangle |$, which has been illustrated on a logarithmic scale in Figs. \ref{fig:amp_scaling}(c) and (d). In the main panel of Fig. \ref{fig:amp_scaling}(c) we have plotted the amplitude as a function of $\ell_1$ for a fixed $\ell_2$ and four different choices of $\ell_y$. These different choices of $\ell_y$ are labeled with the same labels as those in Fig. \ref{fig:amp_scaling}(a). A similar behavior is observed when $\ell_2$ is varied and $\ell_1$ is held constant. For all $\ell_y$ we observe an exponential decay with $\ell_1$ and $\ell_2$. However, we notice that the slope of these lines decreases when $\ell_y$ is increased. In the inset of this subplot, we have plotted the product $\gamma \ell_y$, where $\gamma \equiv \frac{ \partial}{\partial \ell_y} \log|\langle \mathbb{S}_{1,3} V_{R_1\cup R_2}\rangle|$. From here, we notice that the variation in $\ell_y \gamma$ compared to its mean value is less than $2\%$ and therefore negligible. Consequently, the leading order dependence of $\log|\langle\mathbb{S}_{1,3}  V_{R_1\cup R_2}\rangle|$ on $\ell_1$ and $\ell_2$ can be described by $\log|\langle \mathbb{S}_{1,3}V_{R_1\cup R_2} \rangle| \simeq -c_1(\ell_1 + \ell_2)/\ell_y$, where $c_1$ is a positive constant. 

In Fig. \ref{fig:amp_scaling}(d), we have investigated the dependence of $\log|\langle \mathbb{S}_{1,3} V_{R_1\cup R_2}\rangle|$ on $\ell_y$, holding $\ell_1$ and $\ell_2$ fixed. While the asymptotic exponential decay of $\log|\langle \mathbb{S}_{1,3} V_{R_1\cup R_2}\rangle|$ with $\ell_y$ resembles the same exponential decay we observed in Fig. \ref{fig:amp_scaling}(b), the initial increasing behavior is due to $V$ and the fact that increasing $\ell_y$ suppresses the exponential decay of $\log|\langle \mathbb{S}_{1,3} V_{R_1\cup R_2}\rangle|$ in terms of $\ell_1$ and $\ell_2$. Hence, we see that the behavior of $\log|\langle \mathbb{S}_{1,3} V_{R_1\cup R_2}\rangle|$, in the leading orders can be described as the interplay of the SWAP and the exponentiated polarization operator $V$. 

To verify this hypothesis, we have fitted a polynomial to Fig. \ref{fig:amp_scaling}(d) by only considering the inverse $\ell_y$ dependence observed in Fig. \ref{fig:amp_scaling}(c), a constant term, and a linear exponential decay as observed in Fig. \ref{fig:amp_scaling}(c). More specifically, we consider the following fit:
\begin{align}
\log|\langle \mathbb{S}_{1,3} V_{R_1\cup R_2}\rangle| = - c_1(\ell_1 + \ell_2)/\ell_y - c_2 - c_3 \ell_y,
\end{align}
where $c_i$'s are some positive constants which provide an optimal fit for the original curve. The term linear in $\ell_y$ can be understood from the exponential suppression due to the SWAP operator, as discussed above.
The first term proportional to $(\ell_1 + \ell_2)/\ell_y$ can heuristically be associated with the contribution of the excited states when the exponentiated polarization operator $V_{R_1\cup R_2}$ is applied to the ground state. As demonstrated by Resta \cite{Resta1998Quantum}, these contributions in the leading order scale with $\ell_y^{-1}$ when the operator $\prod_{x,y} e^{i 2\pi y \hat{n}/\ell_y}$ is applied to the ground state. Furthermore, since the corresponding Wilson operator $V_{R_1\cup R_2}$ only has support in regions $R_1$ and $R_2$, we expect the $\ell_y^{-1}$ correction to also be extensive, scaling as $(\ell_1 + \ell_2)$.
The optimal fit based on this polynomial is plotted with a red dashed line in Fig. \ref{fig:amp_scaling}(d).

\section{Discussion}\label{secDiscussion}

We have shown how a single wave function can be used to extract the many-body Chern number for both integer and fractional quantum Hall states. Remarkably, the numerical results indicate that a single wave function (or reduced density matrix) on a disk-like open patch of the system is sufficient to obtain the Chern number.

In the FQH case, our formulas require knowledge of an additional topological invariant, $s$, which corresponds to the minimal number of flux quanta that must be inserted into the system to get a topologically trivial excitation. We have discussed how $s$ can be obtained given access to the degenerate ground states on a torus (without twisting the boundary conditions). It remains a fundamental question whether $s$, and other aspects of the intrinsic topological order, can be obtained from a single wave function.

The key point that makes the approach work is twofold: (1) the swap operation gives us access to the space-time path integral where there is non-trivial topology involving the time direction as well, and (2) the two new intersecting non-contractible cycles that are introduced in the space-time manifold due to the swap operation are both normal to the real time direction; therefore, the symmetry defects can be implemented by operators at a fixed real time slice and only require knowledge of the density operator.

We have provided derivations using TQFT for the case of a wave function defined on a cylinder, however it is an assumption that the expectation values of the full microscopic theory map to their counterparts in the TQFT. It is clearly an important open direction then to obtain a completely rigorous derivation of the formulas presented here, without relying on such an assumption. 

Surprisingly, we observe that when symmetry defects are open sheets, e.g., rectangles, our numerical results produce the correct MBCN, while the TQFT prediction is ill-defined and requires regularization, and therefore, it is not a straightforward generalization. Furthermore, there are subtleties related to changing support of the symmetry defect sheets for V and W separately, which could be the subject of further studies.

While our formulas do not require knowledge of the Hamiltonian, they do require knowledge of the time-independent $U(1)$ conserved density operator, $\hat{n}(x,y)$. One could consider instead obtaining the Chern number given the current density operator $\hat{j}_i$ for $i = x,y$. However there is a fundamental difference between $\hat{n}$ and $\hat{j}_i$. $\hat{n}(x,y)$ is distinguished as it generates the symmetry operators on the Hilbert space. Therefore in principle $\hat{n}(x,y)$ can be extracted given the wave function, without any knowledge of the Hamiltonian. For a wave function with $U(1)$ symmetry $|\Psi\rangle $, we can in principle search for $\hat{n}(x,y)$ by considering operators of the form $U = \prod_{(x,y)} e^{i \alpha \hat{O}(x,y)}$. The choice of $\hat{O}(x,y)$ that keeps the wave function invariant for any $\alpha$ then gives the conserved number density $\hat{n}(x,y)$. The current density $\hat{j}_i$, on the other hand, necessarily requires knowledge of the kinetic term in the Hamiltonian, but not the interaction term. It is an interesting question to study how to extract the Chern number given knowledge only of the current density $\hat{j}_i(x,y)$ and a single ground state, but not the full Hamiltonian.

In this paper we have studied ground state wave functions of gapped Hamiltonians, yet there are also situations in which systems with gapped charged excitations and a quantized many-body Chern number can possess gapless neutral modes, such as in the exciton condensate at total filling $\nu_{T} = 1$ \cite{Eisenstein2014Exciton} and the proposed topological exciton metal states in quantum Hall bilayers \cite{Barkeshli2018Topological}. It would be interesting to study whether the approach presented here can also be used to extract the Chern numbers in such gapless states.

\section{Acknowledgements}

We thank Kevin Walker, Hassan Shapourian, Andreas Elben, Beno\^{i}t Vermersch, Guanyu Zhu and Peter Zoller for helpful conversations. ZC, HD, and MH were supported by AFOSR FA9550-16-1-0323 and FA9550-18-10161, ARO W911NF-15-1-0397 and Google AI. HD, MH thanks the hospitality of the Kavli Institute for Theoretical Physics, supported by NSF PHY-1748958. MB is supported by NSF CAREER (DMR- 1753240) and Alfred P. Sloan Research Fellowship. All authors acknowledge the support of NSF Physics Frontier Center at the Joint Quantum Institute. 

\appendix
\begin{widetext}

\section{Proof that $s = \text{max}_i  m_i = \text{lcm}(\{m_i\})$}
\label{zlcm}

Let us label the $M$ topologically distinct anyons in the theory as $\sigma_i \times v^\alpha$, for integer $i= 1,\cdots, k$ and $\alpha = 1,\cdots, m_i-1$. Here $v$ denotes the vison. 
We can correspondingly label the states on a torus as $|\alpha,i\rangle$, which denotes the state where the topological charge as measured along the longitudinal cycle is $\sigma_i \times v^\alpha$. Let $\sigma_1 = 1$ denote trivial topological charge. 

$m_i$ is defined by the minimal integer such that
\begin{align}
\sigma_i \times v^{m_i} = \sigma_i . 
\end{align}
Therefore, any $a$ that satisfies $\sigma_i \times v^a = \sigma_i$ must be an integer multiple of $m_i$.

Furthermore, $s$ is defined by the minimal integer such that
\begin{align}
v^s = 1,
\end{align}
so that $s = m_1$. It follows that
\begin{align}
\sigma_i v^s = \sigma_i ,
\end{align}
which implies that $s$ is an integer multiple of each $m_i$, and furthermore $s = m_1$. Finally, it is clear that $m_i \leq s$. Therefore
\begin{align}
s = \text{lcm} (\{m_i\}) = \text{max}_i m_i. 
\end{align}

\section{Quasiparticle charges are integer multiples of $1/s$}
\label{qpChargeS}

 The electric charge $Q_{a}$ of an anyon $a$ is determined by the mutual statistics $M_{a, v}$ between the vison $v$ and $a$ through
\begin{equation}
e^{2\pi i  Q_{i, \alpha}} = M_{a, v} .
\end{equation}
Thus
\begin{align}
e^{2\pi i  s Q_{a}} = M^s_{a, v} = M_{a, v^s} = 1.
  \end{align}
It follows that $Q_a$ is an integer multiple of $1/s$.

   \section{Proof that $R_y^s | \Psi \rangle \propto |\Psi \rangle$}
   \label{RysPsi}

Here we show that for any state on a torus, $|\Psi\rangle$, $R_y^s |\Psi \rangle \propto |\Psi\rangle$, up to an overall phase factor. Note that we consider the thermodynamic limit where the action of $R_y$ and its  integer powers to a given ground state, will be limited to the ground state space which is demonstrated in \cite{Resta1998Quantum}. 

Since the system is an insulator, we can always diagonalize $R_y$ within the ground state subspace and the eigenvalues should be pure phases in the thermodynamic limit. Let us fix a reference state
\begin{equation}
R_y |1\rangle = e^{i\lambda} |1\rangle,
\end{equation}
where $\lambda$ is a real number.

If we consider a system in the cylindrical geometry with the $x$ direction compactified, the eigenstates of $R_y$ can be obtained by taking a quasiparticle $a$ from one end and moving it to the other end of a cylinder. It can be implemented by the action of a string operator. This defines the state
\begin{equation}
|a \rangle = W_{a} |1 \rangle,
\end{equation}
where $W_{a}$ is the string operator that moves the quasiparticle $a$ along the $y$ direction from one end of the cylinder to the other. The operator $W_{a}$ changes the polarization of the state along the $y$ direction because it creates charge $-Q_{a}$ on one end and $+Q_{a}$ on the other end. 

Therefore the eigenvalues of $R_y$ should be
\begin{equation}
R_y |a \rangle = e^{i\lambda} e^{i2\pi Q_{a}} |a \rangle .
\end{equation}
Since $Q_a$ are all integer multiples of $1/s$, it follows that 
\begin{equation}
R^s_y |a \rangle = e^{is\lambda} |a \rangle .
\end{equation}
Thus for any arbitrary linear combination of the degenerate ground states $|\Psi \rangle = \sum_{a} c_{a} | a\rangle$,
\begin{equation}
    R_y^s |\Psi \rangle = e^{is\lambda} |\Psi \rangle .
\end{equation}

To corroborate the above result through numerical simulations, we numerically simulate the bosonic Pfaffian state with filling fraction $\nu = 1$ on a square torus with $L_x = L_y = 4$. The ground state exhibits three-fold degeneracy. The three ground state are labeled as $\ket{I}$, $\ket{\psi}$ and $\ket{\sigma}$ respectively. The absolute value of $\langle R_y^k \rangle$ is summarized in the following table :

\begin{center}
\begin{tabular}{| c | c | c |} 
 \hline
  Expectation value & $k = 1$ & $k = 2$  \\ [0.5ex] 
 \hline\hline
 $|\bra{I} R_y^k \ket{I}|$ & $5.3\times 10^{-5}$ & $2.93\times 10^{-2}$  \\ 
 \hline
 $|\bra{\psi} R_y^k \ket{\psi}|$ & $5.3\times 10^{-5} $& $ 2.93\times 10^{-2}$ \\
 \hline
 $|\bra{\sigma} R_y^k \ket{\sigma}|$ & $2.35\times 10^{-1}$ & $2.93\times 10^{-2}$  \\
 \hline 
\end{tabular}    
\end{center}

We observe that for the state $\ket{I}$ and $\ket{\psi}$, the absolute value of the expectation value $\langle R_y^k \rangle$ is significantly higher when we choose the power $k = s$. The deviation from unity for $|\bra{\sigma} R_y \ket{\sigma}|$, $|\bra{\sigma} R_y^2 \ket{\sigma}|$, $|\bra{\psi} R_y^2 \ket{\psi}|$, and $|\bra{I} R_y^2 \ket{I}|$ is expected from finite size effects. 

\end{widetext}

\end{document}